%% file: ms.tex
\documentclass[apj,graphicx]{emulateapj}
\usepackage{apjfonts}

\usepackage{graphicx}

\newcommand\suz{{\it Suzaku}}
\newcommand\xmm{{\it XMM-Newton}}

\newcommand\xte{{\it RXTE}}
\newcommand\exo{{\it EXOSAT}}
\newcommand\sax{{\it BeppoSAX}}

\shortauthors{Cackett et al.}
\shorttitle{Archival neutron star Fe lines}

\begin{document}

\title{A comparison of broad iron emission lines in archival data of \\ neutron
star low-mass X-ray binaries}

\author{Edward~M.~Cackett\altaffilmark{1,2}}
\author{Jon~M.~Miller\altaffilmark{3}}
\author{Rubens~C.~Reis\altaffilmark{2,3}}
\author{Andrew~C.~Fabian\altaffilmark{2}}
\author{Didier~Barret\altaffilmark{4}}

\email{ecackett@wayne.edu}

\affil{\altaffilmark{1}Department of Physics \& Astronomy, Wayne State University, 666 W. Hancock St, Detroit, MI 48201}
\affil{\altaffilmark{2}Institute of Astronomy, University of Cambridge,
Madingley Rd, Cambridge, CB3 0HA, UK}
\affil{\altaffilmark{3}Department of Astronomy, University of Michigan, 500
Church St, Ann Arbor, MI 48109-1042, USA}
\affil{\altaffilmark{4}Institut de Recherche en Astrophysique et Planétologie,
9, Avenue du Colonel Roche, BP 44346, 31028 Toulouse Cedex 4, France}

\begin{abstract}
Relativistic X-ray disk-lines have been found in multiple neutron
star low-mass X-ray binaries, in close analogy with black holes across the
mass-scale. These lines have tremendous diagnostic power and have been used to
constrain stellar radii and magnetic fields, often finding values that are
consistent with independent timing techniques. Here, we compare
CCD-based data from \suz\ with Fe K line profiles from archival data taken with
gas-based spectrometers. In general, we find good consistency between
the gas-based line profiles from \exo, \sax\ and \xte\ and the CCD data from
\suz, demonstrating that the broad profiles seen
are intrinsic to the line and not broad due to instrumental issues. However, we do
find that when fitting with a Gaussian line profile, the width of the Gaussian can
depend on the continuum model in instruments with low spectral resolution, though when the different models fit equally well the line widths generally agree. We also
demonstrate that three \sax\ observations show evidence for asymmetric lines,
with a relativistic disk-line model providing a significantly better fit than a
Gaussian.  We test this by using the posterior predictive p-value
method, and bootstrapping of the spectra to show that such deviations from a
Gaussian are unlikely to be
observed by chance.

\end{abstract}

\section{Introduction}

Broad iron emission lines are known to be prominent in the X-ray
spectra of many accreting objects, especially in stellar-mass and
supermassive black holes \citep[for a review, see][]{miller07}.  The
highest quality X-ray observations suggest that a clear asymmetry is
seen in the line profile, as would be expected if the lines originate
from the innermost region of the accretion disk and therefore subject
to strong relativistic effects.  An origin in the innermost,
relativistic regions of the accretion disk is the favored explanation
for the broad lines seen in black hole X-ray binaries and AGN
\citep[for example, see reviews by][ and references
  therein]{fabian00,reynolds03,miller07}.  This interpretation has
held up to many alternative models for line broadening and the nature
of the innermost accretion flow
\citep[e.g.,][]{reynoldswilms00,reynolds_mcg6_09,fabian09,miller10}.

In neutron star low-mass X-ray binaries (LMXBs), iron emission lines
were first discovered in the mid-1980s with \exo{} and {\it Tenma}
\citep{white85,white86,hirano87} and have been observed by every other
major X-ray mission since (e.g. {\it RXTE, ASCA, BeppoSAX, Chandra,
  XMM-Newton} and \suz).  Compared to the lines in black hole X-ray
binaries and AGN, the lines seen in neutron star LMXBs are typically
much weaker, with a peak deviation from the continuum of less than
10\%.  This likely hindered the detection of relativistic lines in
neutron stars, which are otherwise expected since the potential around
a neutron star is somewhat similar to that around a black hole with a low spin
parameter.

Recently, sensitive CCD-based spectroscopy with {\it Suzaku} and
{\it XMM-Newton} has started to probe the line shape in neutron star
LMXBs \citep{bhattacharyya07,cackett08,cackett_j1808_09,cackett10,papitto09,
reis09_1705,dai09,dai10,disalvo09,iaria09,egron11}.  In many cases,
relativistic line shapes are found with inner disk radii, measured using
relativistic line models, close to expectations for the stellar radius.
Moreover, neutron star magnetic field strength estimates based on the inner disk
radius for two accreting pulsars are consistent with measurements
from timing \citep{cackett_j1808_09,miller11,papitto09,papitto11}. 
Gas spectrometer observations have also revealed relativistic lines in 4U~1705$-$44 \citep{piraino07, lin10}.

For bright sources, photon pile-up can affect CCD spectra
\citep{ballet99,davis01}.  Pile-up occurs when more than one photon
arrives in an event box in a single detector frame time.  Thus,
rather than recording the energy and time of each individual photon,
all photons arriving at that location within a single frame time get
counted as a single photon, with an energy equal to the sum of the
energy of the individual photons \citep[see][ for a more detailed
  discussion of this effect]{davis01,miller10}.  Therefore, the shape
of the detected spectrum is altered, with the spectrum becoming
artificially stronger at higher energies and weaker at lower energies.
This could, of course, alter the ability to accurately recover the
iron K emission line profile in observations strongly affected by this
process.  In order to address this issue,
\citet{miller10} performed a wide range of simulations based on
typical black hole and neutron star LMXB spectra.  These authors found
that pile-up does not broaden the line profiles recovered, and in the
most extreme cases it acts in the opposite sense, leading to a
narrower line profile.

An analysis of \xmm\ spectra of neutron star LMXBs by \citet{ng10}
came to an opposite conclusion about how pile-up affects the line
profiles.  These authors looked at \xmm\ spectra taken in pn timing
mode, where the detector is read out continuously, and found that
pile-up was significant in many of the observations.  To mitigate
these effects, they extracted spectra only from the wings of the
point-spread function (PSF), excluding the columns on the detector
where the count rate was highest and thus has the most severe
problems.  They find that the resulting spectra are consistent with a
Gaussian line profile, and that relativistic, asymmetric line models
are not required to fit the data.  These results differ from those
obtained with {\it Suzaku}, wherein excising successive annular
regions of the PSF yields highly consistent and relativistic line
profiles \citep{cackett10}.  Although \citet{ng10} note that continuum
modeling, background subtraction, charge transfer inefficiencies, and
X-ray loading of offset maps (an instrumental effect that is unique to
the EPIC-pn camera) have a significant effect on the overall line
profile, \citet{ng10} conclude that photon pile-up falsely leads
to relativistic line profiles.  However, in the spectra excluding the core of the PSF, the quality of the spectra are severely limited as the vast majority of source counts are excluded.

In order to place new results from CCD spectroscopy of Fe K lines in
neutron stars into a broader context, and in order to better
understand the results of data from CCD and gas spectrometers, we
systematically compare data from three gas spectrometer missions that
are unaffected by pile-up (\exo, \xte, \sax) with CCD data from \suz.
We choose to compare with the \suz\ data, because, as discussed in
length by \citet{miller10}, pile-up is much less of a problem with the
\suz/XIS detectors than with \xmm/pn -- the much broader PSF of
\suz\ means that the source flux is spread over many more pixels.  As
shown by \citet{cackett10}, extracting the source spectrum from an
annulus, excluding the central core of the \suz\ PSF, does not change the
shape of the line profile obtained.  In this paper, we show that the gas
spectrometer data give good consistency between the different missions
and with the CCD \suz\ data, supporting the relativistic
interpretation of the iron line origin.

\section{Data sample and reduction}

In this paper, we analyze data from four different missions: \exo, \xte,
\sax, and \suz.  The detectors we utilize from \exo, \xte\ and \sax\ are gas
spectrometers, whereas the \suz\ detectors covering the Fe K line region are
CCDs. The combination of both effective area and energy resolution is
important for
iron line spectroscopy and for the discussion of our results.  The parameters
for the detectors we use follows here.  The \exo/GSPC has an energy resolution
(FWHM) of approximately 10\% at 6 keV \citep[$\sim$0.6 keV][]{peacock81} and
an effective area of around 150 cm$^2$ at 6 keV.  The \sax/MECS instrument is
very comparable to \exo/GSPC, with a similar energy resolution ($\sim$0.5 keV)
and effective area at 6 keV \citep{boella97}.  The proportional counters on
\xte, however, have
a significantly higher effective area (about 1300 cm$^2$ per PCU), but a
significantly lower energy resolution of approximately FWHM $\sim$1 keV at 6
keV \citep{jahoda96}.  By comparison the two working front-illuminated XIS
detectors on \suz\
have a combined effective area of around 600 cm$^2$ at 6 keV, and by far the
best energy resolution of the detectors considered here, at approximately 0.13
keV \citep{mitsuda07}.  

Broadband energy coverage is also important in determining the shape of the
continuum either side of the iron line region.  The energy range of \exo/GSPC
is dependent on the gain used during the observation.  Most of the observations
considered here are restricted to the 2 -- 16 keV range (Gain 1), though some
extend to higher energies (Gain 2), but in almost all those cases the S/N
drops
sharply above 20 keV.  For \xte\ we analyze data from the PCA only, which
typically gives good spectra in the range from 3 - 30 keV.  \sax\ has several
detectors to provide a broad energy coverage.  Here, we used the LECS (0.5 --
3.5 keV), MECS (2 -- 10 keV) and PDS (15 -- 200 keV, with the upper limit
dependent on S/N) to provide broad energy coverage.  In the case of \suz, we fit
the XIS detectors in the energy range 1 -- 10 keV, and use the PIN hard X-ray detector
to provide broad band coverage from 12 -- 50 keV (with the upper limit dependent
on S/N).

We restrict our choice of sources here to four sources that have consistently
shown the strongest broad iron emission lines in multiple observations by
different missions, namely Serpens X-1, GX 349+2, GX 17+2 and 4U~1705$-$44.   We
analyze every \exo\ and \sax\ observations of these sources.  For \suz\, we use
the \suz\ spectra analyzed by \citet{cackett10} here, please see that paper
for details of the data reduction of those data.  With \xte\ many observations
of each source exist.  As the focus of this work is on a comparison of
the line profiles from different missions, as opposed to a detailed study of the
line variability in each of the sources, we choose to look at only 3 \xte\
spectra of each source.  The specific spectra were initially chosen to be the three
single longest continuous spectra of each source, but to ensure that we sampled a range of source states, we chose other observations in some cases.  See Table~\ref{tab:obs} for details of the observations analyzed here.

\input{tab1.tex}

For \exo, and \sax\ we used the pipeline produced spectra available from the
HEASARC database.  The \exo/GSPC pipeline produced spectra are time-averaged, background subtracted spectra.  For bursting sources, the spectra have bursts already removed.  The pipeline produced spectra are provided along with the corresponding response matrices.  Further details of the \exo{} data archive are given on the HEASARC website\footnote{http://heasarc.gsfc.nasa.gov/docs/journal/exosat\_archive6.html}.  The \sax\ pipeline spectra for the LECS, MECS and PDS are used.  The PDS data are already background-subtracted, for the LECS and MECS we use the appropriate background file for the 8 arcmin extraction radius used for the source spectra, along with the appropriate response matrices from September 1997.  For \suz, see the extensive details given in
\citet{cackett10}, here we use the spectra from that paper directly.  Finally, for \xte, we use spectra from PCU2 only to provide the most reliable spectrum, with the data reduced following the standard procedures.  Briefly, the standard `goodtime' and `deadtime' corrections are applied, selecting data when the pointing offset $<0.02^\circ$ and the Earth-limb elevation angle was $>10^\circ$.  Spectra from PCU2 were extracted from the Standard 2 mode data, and 0.6\% systematic errors were applied to every channel.  The appropriate background model for bright sources was used and response matrices were created with the PCARSP tool.

\section{Spectral fitting and analysis}

Spectral fitting is performed with XSPEC v12 \citep{arnaud96}, and
uncertainties are quoted at the 1$\sigma$ level of confidence throughout.

Broadband spectral fitting of neutron star LMXBs is often degenerate - a
combination of several different models usually fit the data well
\citep[e.g.,][]{barret01}. This is a well known problem that led to the Eastern
\citep{mitsuda89} versus Western \citep{white88} model debate. Here, we
choose to
explore two widely used continuum models for neutron star LMXBs in order to
assess the importance of continuum choice on the iron emission line profiles. We
first use a model consisting of a single-temperature blackbody, disk blackbody
and power-law (if needed), which we have successfully used in the past
\citep[e.g.,][]{cackett08,cackett10} and is based on the results of testing
multiple
continuum models by \citep{lin07}.  We refer to this model as Model 1. The
second model we test consists of a single-temperature blackbody, Comptonized
component (comptt in xspec), and in a number of cases an additional power-law
(if needed). Hereafter, we refer to this model as Model 2. This model is widely
used in the literature, for instance, see
\citet{barret01,disalvo00,disalvo01,iaria04}.

In both models
Galactic photoelectric absorption is included with the phabs model.  We mostly
leave this as a free parameter in the model.  However, with the \xte\ data we
always fix the $N_{\rm H}$ parameter at the \citet{dickey90} values (using the
HEASARC $N_{\rm H}$ tool),
except for GX~17+2 where we use the $N_{\rm H}$ value determined from fitting
the neutral absorption edges present {\it Chandra} gratings data
\citep{wroblewski08}, which is found to be higher than the \citet{dickey90}
value. For spectra from the other missions, only when the $N_{\rm H}$ parameters
tends to 0, or becomes significantly higher than the values in the literature do
we fix the $N_{\rm H}$.

Initially, we fit the iron line with a Gaussian.  All parameters (line energy,
width and normalization) are free parameters in the fits, though the line energy
of the Gaussian is restricted to be within the 6.4 -- 6.97 keV range.

Brief notes on individual observations follow.  Please see Cackett et al.
(2010) for notes on the \suz\ data.

\subsection{\exo}

{\it Ser~X-1:}  The spectrum is fit from 2 - 16 keV.  For the fits with Model 2
we find an $N_{\rm H}$ significantly higher than the Dickey \& Lockman (1990)
value, thus for Model 2 we fix $N_{\rm H} = 4\times10^{21}$ cm$^{-2}$.  This is similar to the value found when fitting {\it XMM-Newton} spectra of Ser~X-1 \citep{bhattacharyya07,cackett10}.

{\it GX~349+2:}  For all three observations we fit the data from 2 - 20 keV
(above 20 keV the S/N drops quickly).

{\it GX~17+2:}  We find for all observations that the spectrum turns up below 3
keV leading to too low an $N_{\rm H}$ (tending to zero). We therefore ignore below 3 keV and fix
$N_{\rm H} = 2.38\times10^{22}$ cm$^{-2}$ \citep{wroblewski08}, this value is consistent with what we find when fitting the \sax\ and \suz\ observations of GX~17+2.  The final
observation (ObsID: 75122) has very poor calibration and there is a strong
feature at 4.5 keV and between 10 -- 14 keV (likely caused by
variations in intrinsic detector gain, a known issue with \exo). We therefore
do not analyze that
dataset here.  For ObsID 33715 and 33781 we fit between 3 -- 16 keV.  For ObsID
58809 and 60698 a different gain was used, thus we fit from 3 -- 20 keV (above
20 keV the S/N drops).

\begin{figure*}
\centering
\includegraphics[angle=270, width=8cm]{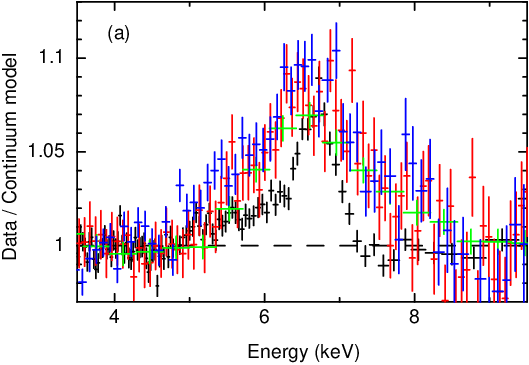}
\includegraphics[angle=270, width=8cm]{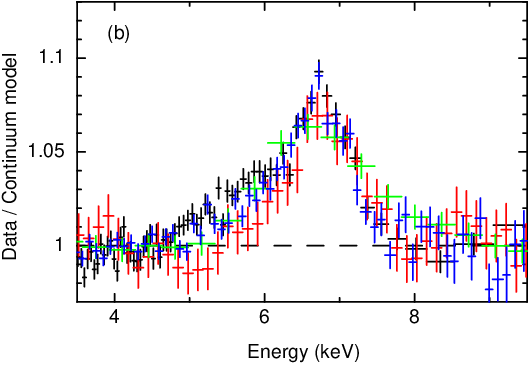}
\includegraphics[angle=270, width=8cm]{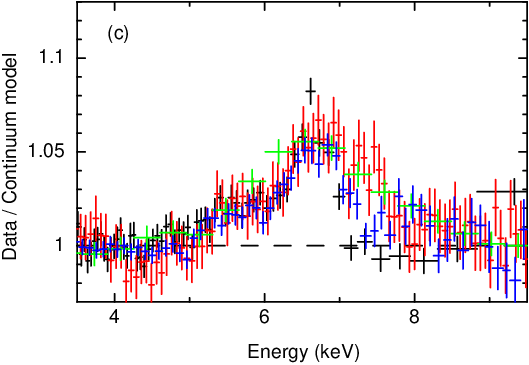}
\includegraphics[angle=270, width=8cm]{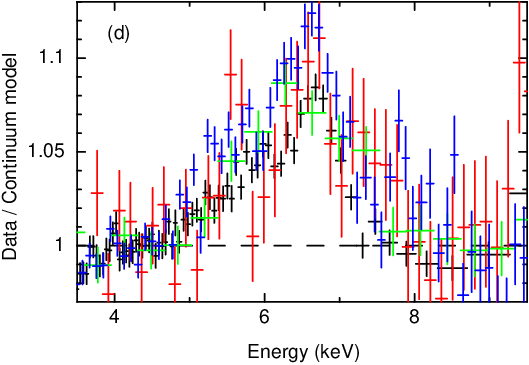}
\caption{A comparison of iron emission line profiles from the four different
missions considered here, and good agreement is seen in most cases.  Panel (a)
shows Serpens X-1, (b) GX~349+2, (c) GX~17+2, and (d) 4U~1705$-$44.  In all
panels, black is \suz/XIS (all FI combined), red is \exo/GSPC, blue is
\sax/MECS, and green is \xte/PCA.}
\label{fig:comparison}
\end{figure*}

{\it 4U 1705$-$44:}  We find no clear line detection in any of the
observations.  Upper limits on the line intensity are comparable or greater than the line intensities seen with other missions, indicating that the spectra do not have the sensitivity to detect the Fe line.  The strongest evidence for a line is in ObsID 60407 which is shown
in Figure~\ref{fig:comparison} (bottom right hand panel, red). However, the
Gaussian component in this observation is required at less than the 2$\sigma$
level. We therefore do not consider these data any further here.

\subsection{\xte}

{\it Ser~X-1:}  We fit the spectra from 3 -- 20 keV.

{\it GX~349+2:}  The spectra are fit from 3 -- 30 keV.

{\it GX~17+2:}  The spectra are again fit from 3 -- 30 keV.

{\it 4U~1705$-$44:}  The spectra are fit from 3-25 keV for observation 2 and 3.  For observation 1, where the source is in a hard state, we fit from 3 - 50 keV.

\subsection{\sax}

We fit the \sax\ spectra allowing an offset between the LECS, MECS and
PDS spectra.  The constant between LECS and MECS was constrained to be within
0.7 - 1.0, and the constant between the PDS and MECS was constrained to be
within 0.7 - 0.95.  Unless otherwise stated, we fit the LECS between 1 -- 3.5
keV, the MECS from 2 -- 10 keV and the PDS from 15 keV with the upper limit
determined by the quality of the data at higher energies.  As noted below, in several of the spectra, there is a feature at $\sim2.6$ keV.  This feature as been discussed in previous analyses of these data by \citet{iaria04} and \citet{piraino07}, who fit the feature and discuss its potential origin.  While the feature may be real, it occurs at a region in the spectrum of both the LECS and MECS where there is a significant change in the response of both detectors, and thus potentially instrumental.  When we fit the feature with a Gaussian, we get parameters for this line that are consistent with the previous analysis of \citet{iaria04} and \citet{piraino07}.  Here, we choose to ignore this section of the data in the few cases where it is present, as it is not the focus of this work.  This approach, or the approach of fitting it with a Gaussian lead to consistent spectral fits at better than the 1$\sigma$ level.   We generally only fit the LECS data  above 1 keV as we often found significantly reduced signal-to-noise there.

{\it Ser~X-1:}  Here, the LECS data below 1 keV is of good quality,
thus we extend the spectral fitting range from 0.5 -- 3.5 keV.  We fit the PDS
from 15 -- 30 keV.

{\it GX 349+2:}  We find a poor reduced chi-squared when fitting the standard
MECS range due to a strong feature at around 2.6 keV in all 3 spectra.  We
therefore fit the MECS data from 3-10 keV in all cases.  For the PDS data we fit the
range 15 -- 90 keV for observation 1 which has a strong hard tail
\citep{disalvo00}, where as
for observations 2 and 3 where this hard tail is not significant
\citep{iaria04}, we fit from 15
-- 40 keV. We did not achieve any fits with
$\chi^2_{\nu} < 2$ when using Model 1. 

{\it GX 17+2:} We fit the PDS from 15 -- 50 keV for all 5 observations.

{\it 4U 1705-44:} In observation 1 we again find significant residuals at
around 2.6 keV in the LECS and MECS spectra.  We therefore fit the LECS data from 1 -- 2.5
keV and the MECS data from 3 -- 10 keV.  We fit the PDS data from 15 -- 50 keV. 
Observation 2 displays a strong hard tail out to 100 keV and thus we fit the PDS data up to
100 keV.  However, an unbroken power-law does not fit the data well, thus we
use an exponentially cut-off power-law for this observation.

\section{Results}
Spectral fitting results are given in Tables \ref{tab:exo_diskbb_bb_gauss} to
\ref{tab:suz_bb_comptt_gauss}, using continuum models 1 and 2 and modeling the iron
line with a Gaussian.  We summarize the resulting line profiles in Figure
\ref{fig:comparison} where we show a comparison of the iron line profiles for
the four different objects and four different missions.  For each object and
mission we show only one spectrum.  The profiles shown here are obtained from
fitting the continuum model excluding the 5 - 8 keV range, and are plotted as
the ratio of the data to the continuum model.  We have chosen the continuum
model that gives the best (lowest $\chi^2$) fit.  The profiles show remarkable
consistency between the different missions.  All observations analyzed here give a positive detection of the iron, though this is no surprise given that these sources were chosen for that specific reason.   We now discuss the main findings
resulting from our analysis.

\input{tab2.tex}
\input{tab3.tex}
\input{tab4.tex}
\input{tab5.tex}
\input{tab6.tex}
\input{tab7.tex}
\input{tab8.tex}
\input{tab9.tex}

\subsection{Asymmetric line profiles}
Our previous analysis of \suz\ data \citep{cackett08,cackett10} has shown
evidence for asymmetric line profiles, which can be fit well by a relativistic
disk line model. The iron lines in \exo\ and \sax\ have been
analyzed in the past
\citep[e.g.,][]{gottwald95,disalvo01,oosterbroek01,iaria04}, however, generally
only a Gaussian line profile has been considered.  An exception to this is the case of 4U~1705$-$44, where both \citet{piraino07} and \citet{lin10} have fitted a relativistic diskline model to \sax\ data, finding that it is a better fit than a simple Gaussian.   Here, we also fit the iron
lines with a relativistic disk-line model \citep[diskline,][]{fabian89}. Of the eight \exo\ spectra we analyze here, six show a
decreased value of $\chi^2$ when using a diskline model rather than
a Gaussian.  However, in all those cases the fit with a Gaussian already
provides a reduced-$\chi^2 < 1$, thus the improvement is not statistically
significant.  The parameters of the diskline fits are reasonable, and consistent with typical parameters \citep[such as in][]{cackett10}, but are not very well constrained.  In the case of \sax, a diskline generally provides equivalently good fits to the data.  However, in three cases, we find that a diskline fits
the data significantly better than a Gaussian.

The three cases where a diskline is a significantly better fit are GX~349+2 observation 1, GX 17+2 observation 1 and 4U~1705$-$44 observation 1.  Note that the 4U~1705$-$44 observation is the same one discussed in detail by \citet{piraino07}, who also find an improvement in $\chi^2$ when using a diskline compared to a Gaussian.
The parameters for the diskline fits to these three spectra are given in
Table~\ref{tab:diskline}.  We give the parameters using the Model 2 continuum
as these provide a better fit in all cases.  A comparison between the diskline fit parameters in Table~\ref{tab:diskline} with previously published fits to these sources show they are consistent.  Our fits to 4U~1705-44 are consistent with similar fits performed by \citet{piraino07} to the same data.  Furthermore,  the diskline parameters we find here are all reasonable when comparing with fits to \suz\ and \xmm\ data of GX 349+2 \citep{cackett08, cackett10, iaria09}, GX 17+2 \citep{cackett10} and 4U 1705-44 \citep{reis09_1705, disalvo09, cackett10, dai10,lin10}.  In particular, the inner disk radius and equivalent width are consistent with this previous work for similar source states.

\input{tab10.tex}

\begin{figure*}
\includegraphics[angle=270,width=17cm]{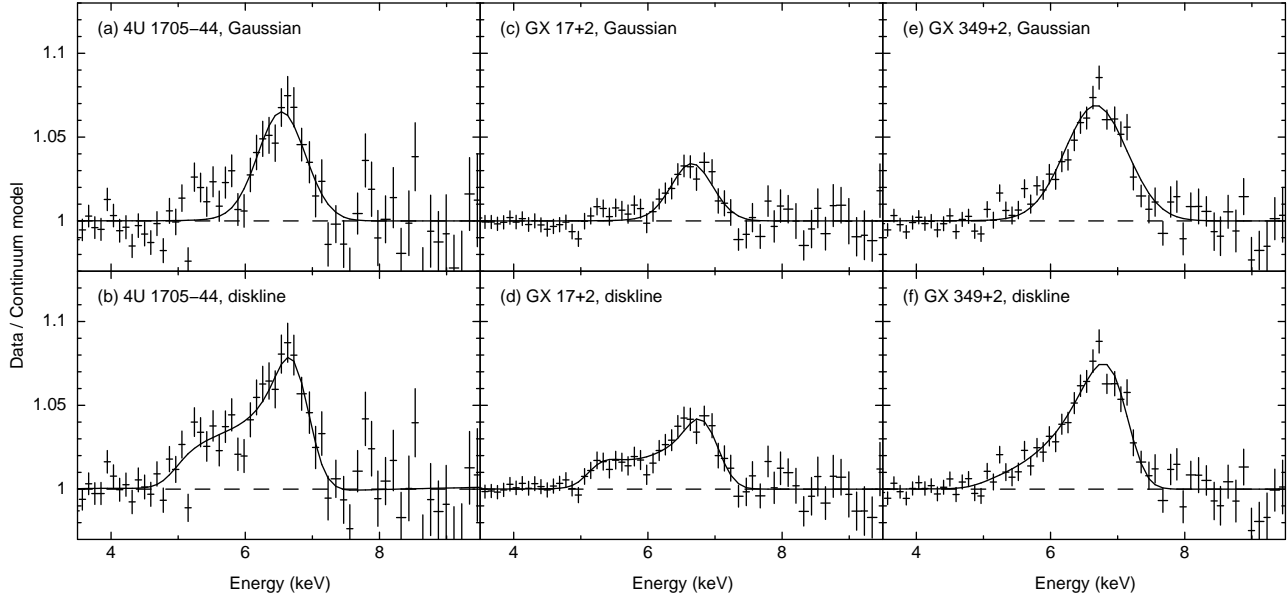}
\centering
\caption{Line profiles from \sax\ observations of 4U~1705+44 (left, observation 1),
GX~17+2 (center,  observation 1) and GX~349+2 (right,  observation 1) fit with both a
Gaussian (top), and diskline (bottom).  In all three cases, the diskline gives a
statistically better fit.}
\label{fig:saxdiskline}
\end{figure*}

In Figure
\ref{fig:saxdiskline} we show the line
profiles with both a Gaussian and diskline model.  The diskline fit to
observation 1 of GX~349+2 gives an improvement of $\Delta\chi^2 = 22.6$ for 2
additional degrees of freedom.  This corresponds to a F-test probability of
$4.1\times10^{-4}$, approximately a 3.5$\sigma$ significance. Observation 1 of
4U~1705$-$44 shows an decrease of $\Delta\chi^2 = 28.2$ when using a diskline
model rather than a Gaussian, which corresponds to an F-test probability of
$6.6\times10^{-5}$, or 4.0$\sigma$ significance.  Finally,  \sax\
observation 1 of GX~17+2 shows a decrease of $\Delta\chi^2 = 14.7$ when using a
diskline model rather than a Gaussian, which corresponds to an F-test
probability of $2.3\times10^{-3}$, or 3.0$\sigma$ significance.  It is important
to also take into account the number of observations we searched to find the
asymmetric lines, as this will reduce the significance.  In total, we
compared diskline and Gaussian fits to a total of 19 observations from \exo\ and
\sax . The probabilities given above should therefore be multiplied by 19,
reducing the significances to 2.7, 3.2, and 2.0 $\sigma$ for GX~349+2, 4U
1705$-$44, and GX~17+2, respectively.

\begin{figure}
\centering
\includegraphics[angle=270,width=8cm]{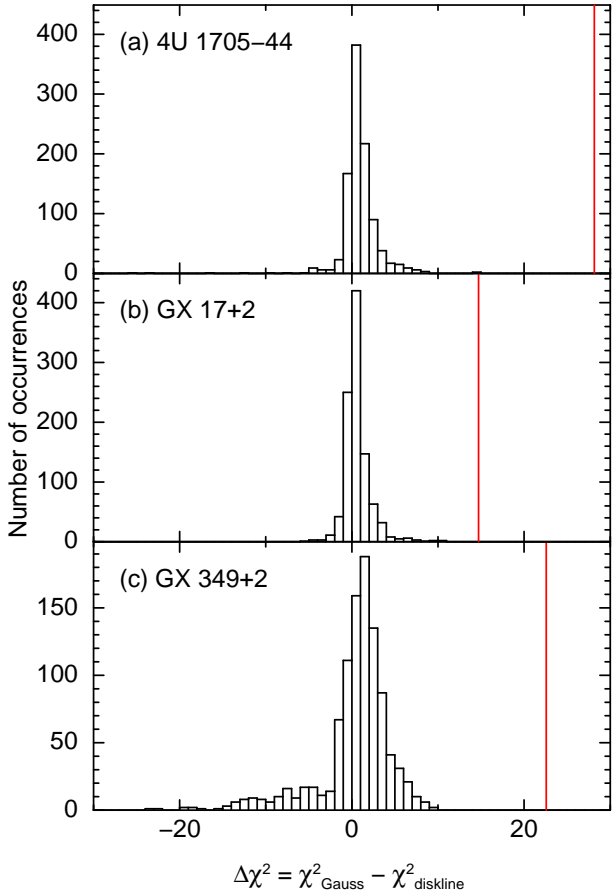}
\caption{Posterior predictive distributions for testing the likelihood that the
diskline model gives the observed $\Delta\chi^2$ by chance for the \sax\
observations of (a) 4U~1705-44, (b) GX~17+2 and (c) GX~349+2 (observation 1 for each object). The distributions
are histograms of $\Delta\chi^2 = \chi^2(Gaussian) - \chi^2(diskline)$ from the
1000 simulations. The red, solid line indicates the $\Delta\chi^2$ measured from
fitting the data.  In none of the cases do any simulations show a
$\Delta\chi^2$ as large as observed, indicating that the diskline model is
strongly preferred.}
\label{fig:ppp}
\end{figure}

\citet{egron11} discuss how under the criteria of \citet{protassov02}, the
F-test can be properly applied to comparing Gaussian and diskline fits.  \citet{egron11} also apply the posterior predictive p-value test \citep[ppp,][]{hurkett08} as a method to determine whether a Gaussian or diskline provides a better fit.  Here, we also apply this ppp test, and secondly we perform bootstrap resampling \citep{efron79} of the spectra to further test the significance of the improvement in using the diskline model instead of a Gaussian.
Essentially, the ppp test involves a
Monte Carlo simulation to test the likelihood that the diskline model gives the
improvement in $\chi^2$ by chance.  First, we find the best-fitting Model 2
using a Gaussian to fit the Fe K line (we use Model 2 as it gives better fits than Model 1).  Next, we simulate 1000 sets of spectra
(LECS, MECS and PDS spectra for each case) with the model parameters
randomly drawn using the covariance matrix of the best-fit and using the
detector responses, background spectra and exposures from the real data in the
simulations. These 1000 simulated spectra are then fit with the same (Gaussian)
model, and also with the diskline model. The posterior predictive
distribution is defined by the values of $\Delta\chi^2$ from comparing fits to
the simulated spectra with the Gaussian and diskline models. The ppp value is
then defined by the number of instances where $\Delta\chi^2$(simulation) $>$
$\Delta\chi^2$(data), see equation 12 and section 3.2 in \citet{hurkett08}.  In
Figure \ref{fig:ppp} we show the posterior predictive distributions for the
three cases (discussed above) where we find a significant improvement using
diskline based on the F-test alone.  In all three cases we do not find a single
simulation with a $\Delta\chi^2$ greater than as measured by the data.  Given
that we find zero instances where $\Delta\chi^2$(simulation) $>$
$\Delta\chi^2$(data) we cannot directly calculate a ppp value.  However, we can infer that the confidence level indicated by the results corresponds to better than 99.9\% level (i.e. less than 1 occurrence in 1000 simulations), which indicates an improvement at better than the 3.29$\sigma$ level.  Thus, the
simulations strongly support the F-test results that a diskline is the preferred model.  

\begin{figure}
\centering
\includegraphics[angle=270,width=8cm]{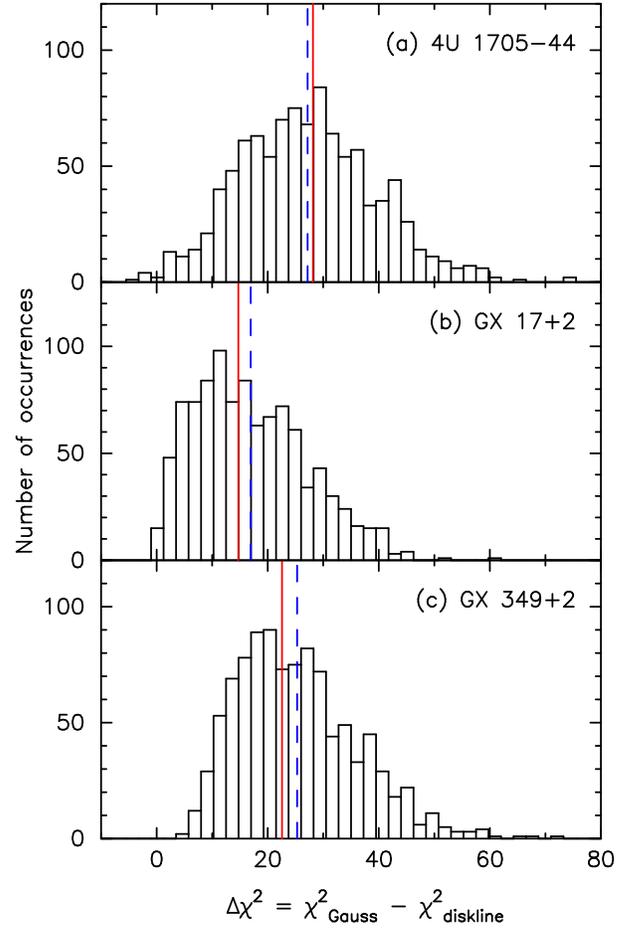}
\caption{Distributions of $\Delta\chi^2$ from bootstrap resampling of the  \sax\
observations of (a) 4U~1705-44, (b) GX~17+2 and (c) GX~349+2 (again, observation 1 for each object). The distributions
are histograms of $\Delta\chi^2 = \chi^2(Gaussian) - \chi^2(diskline)$ from the
1000 resamplings. The red, solid line indicates the $\Delta\chi^2$ measured from
fitting the data. The blue, dashed line indicates the mean $\Delta\chi^2$ from the simulations.}
\label{fig:bootstrap}
\end{figure}

We also perform another test of whether the diskline model fits better than a Gaussian by employing bootstrap resampling \citep[see, e.g.][]{efron79}.  This allows an estimation of the distribution in $\Delta\chi^2$ based on the distribution of the observed data.  Bootstrap resampling consists of resampling the data with replacement.  Thus, applying to spectra with $N$ events, we randomly select $N$ events from the spectrum, with replacement to create a new resampled spectrum.  In this way some events get selected multiple times, where as others will not get selected at all.  Doing this, we create 1000 resampled spectra from both the LECS and MECS detectors.  We cannot do this for the PDS spectrum, as the pipeline produced spectrum is already background subtracted, and thus does not allow for a resampling of the detector events.  However, as the region of interest (the Fe K line) is within the MECS detector, this is not an issue here.  Thus, we resample the LECS and MECS spectra, but for the PDS we use the original data in the fit each time.  For each set of resampled spectra, we fit both the diskline and Gaussian models (Model 2), and measure the $\Delta\chi^2$ between the fits.  In Figure~\ref{fig:bootstrap} we show the distribution of $\Delta\chi^2$ for the 1000 resamples for each three cases.  As should be expected, the mean of the distribution is at approximately the $\Delta\chi^2$ of the data.  Counting the fraction of samplings where $\Delta\chi^2 > 0$ (diskline better than a Gaussian) helps understand the significance of the diskline being a better fit than the Gaussian.  For 4U~1705$-$44 we find 99.4\% of resampled spectra have $\Delta\chi^2 > 0$, for GX~17+2 we find 99.8\% and for GX~349+2 we find 100\%.  Again, strongly supporting that a diskline is the preferred model over a Gaussian.

To determine whether the models used are appropriate, we perform a Kolmogorov-Smirnov (KS) test \citep[e.g.][]{press92}.  In the KS tests we determine the cumulative distribution function of the \sax /MECS spectrum as a function of energy, and compare this with the cumulative distribution function of the model.  We do this for both the model with a Gaussian and the model with a diskline.  We find that for all three cases for both the Gaussian and the diskline model that the KS probability = 1.0, indicating that both models are adequate descriptions of the data.

\begin{figure}
\centering
\includegraphics[angle=270,width=8cm]{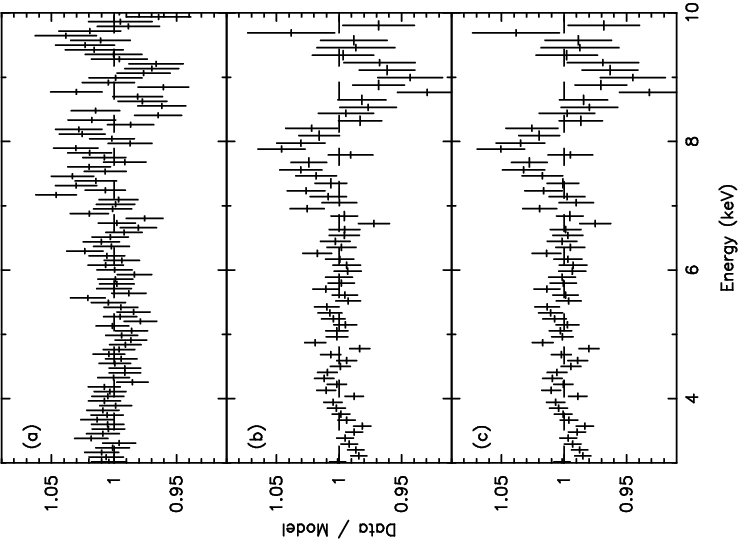}
\caption{Data/model for fits to the (a) \exo\  and (b) \sax\ data of
Ser~X-1.  A diskline model was used to fit the Fe line, with all parameters
except the normalization fixed at those determined from fits to \suz\ data in
\citet{cackett10}.   There are no large residuals in the Fe line region, indicating the
\suz\ diskline model fits well, though some residuals are apparent above 7 keV in the \sax\ data. Panel (c) is for the same data as (b) except that here all the line parameters are allowed to vary.}
\label{fig:diskline_residuals}
\end{figure}

It is also important to consider whether we should have observed more asymmetric
line profiles in these archival data.  Clearly, the spectral resolution of
\xte/PCA is too low to be able to show line shapes similar to the \suz\ line
profiles.  For \exo\ and \sax, where the spectral resolution (FWHM) at
6 keV is 0.6 and 0.5 keV respectively, the answer is not as immediately clear. 
The asymmetric Ser~X-1 line as seen by \suz\ shows a narrow peak and broad wing
which is not apparent in the \exo\ and \sax\ data (see top left panel of
Fig.~\ref{fig:comparison} where the difference between the \suz\ and other line
profiles can be seen). We have fit the \exo\ and \sax\ spectra using a
diskline with the parameters fixed to the \suz\ values \citep{cackett10},
except the line normalization which was variable in order to match the
line strength.  This intrinsically asymmetric line profile fits both spectra
reasonably well ($\chi^2 (dof) = 214.9 (185)$ for \exo\ and $\chi^2 (dof) = 241.1 (135)$ for \sax\ with Model 1), and there are no large residuals in the iron line region, though some residuals are present above 7 keV in the \sax\ data (we show the residuals in Figure~\ref{fig:diskline_residuals}).  If the line parameters are allowed to be free parameters in the fit the parameters are not well constrained for \exo\ (for instance the inclination tends to $90^\circ$).  However, for \sax\ the $\chi^2$ improves slightly to $\chi^2 (dof) = 231.5 (131)$, and apart from the normalization, the parameters remain close to the \suz\ values.  For comparison, the best fitting line parameters from fitting the \sax\ data are: $E = 6.97^{-0.04}$ keV, $\beta = -4.3\pm0.5$, $R_{in} = 6^{+1}$ GM/c$^2$, $i = 28\pm1$, norm = $(9.0\pm0.7)\times10^{-3}$.  The difference between the \suz\ profile of Ser X-1 and the \exo\ and \sax\ profiles is the most strikingly of the four sources, thus the fact that the \suz\ profile still fits the spectrum reasonably well demonstrates the significant smoothing by the lower resolution of those instruments, which may generally mask line asymmetry by blurring out the structure.

Furthermore, we perform additional simulations to see whether the $\Delta\chi^2$ value that we observe is what would be expected if the line profile during the \sax\ observations is the same as during the \suz\ observations.  We therefore take the best-fitting \sax\ continuum (Model 2) and add a diskline with the same parameters as the best \suz\ fits \citep[from][]{cackett10}, with only the normalization changed to match the EW seen in \sax\ observation.  We perform 100 simulations, fitting the simulated spectra with both a Gaussian and diskline, and look at the distribution of $\Delta\chi^2$.  For GX 349+2, we find that the median $\Delta\chi^2 = 31$, comparable to the $\Delta\chi^2 = 22.6$ that we observe from the real data.  For GX 17+2, we  find that the median $\Delta\chi^2 = 0.5$, much less significant than the $\Delta\chi^2 = 14.7$ from the real data.   For 4U 1705-44. we find that the median $\Delta\chi^2 = 18.4$, a little less than the $\Delta\chi^2 = 28.2$ from the real data.  This demonstrates that for GX~349+2 and 4U~1705$-$44 we see approximately the improvement in $\chi^2$ that we would expect based on the \suz\ line profiles, where as for GX~17+2 we see an even better improvement than expected.

\subsection{Continuum model dependence}

We fitted the data with two different continuum models.  Model 1 contained a single temperature blackbody, disk-blackbody and a power-law (when needed), and Model 2 contained a single temperature blackbody, thermal Comptonization and a power-law (when needed). We generally find that Model 2 gives a better fit to the spectra than Model 1, and there is no clear dependence on source state or flux for this.

\begin{figure}
\centering
\includegraphics[angle=270, width=8cm]{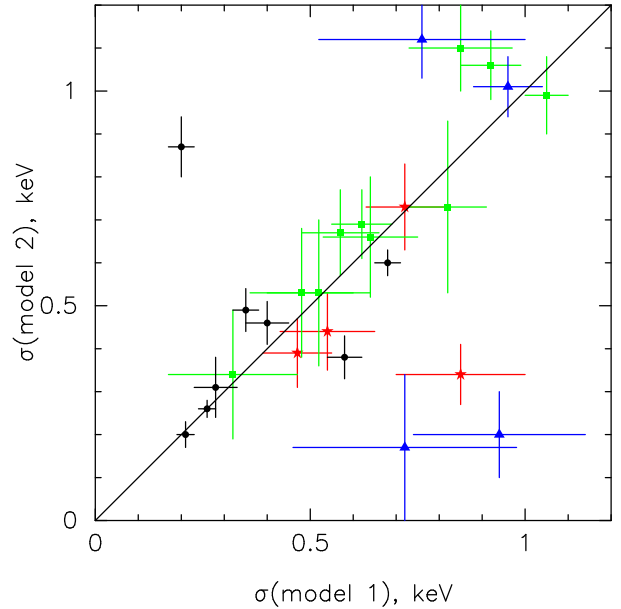}
\caption{A comparison of the Gaussian line widths measured when fitting the
spectra with two different continuum models (model 1: diskbb + blackbody +
power-law or model 2: bbody + comptt + power-law). Black circles represents data
from \suz, red stars are from \exo, blue triangles are from \sax, and green
squares are from \xte. }
\label{fig:sigmas}
\end{figure}

We consider how robust the Gaussian line widths are to the different continuum models.    In Figure \ref{fig:sigmas} we compare the line widths ($\sigma$) obtained when using both continuum models.  We only compare observations where the two different models give equally good fits.  We define equally good fits as the $\Delta\chi^2$ between the two models being less than 11.8.  This is equivalent to 3$\sigma$ for the two additional degrees of freedom of Model 2 compared to Model 1.  The exception to this is the \suz\ data where we find that only one observation has equally good fits, and so for illustrative purposes we show all the \suz\ observations.

Generally, we find that there is a good agreement between the two models when equally good fits are compared.  The biggest outliers that are not in agreement all have large uncertainties in the line widths.  These are generally \exo\ and \sax\ observations -- the two missions with the lowest effective area, and also significantly lower spectral resolution than \suz. Given that most observations
here have similar exposure times, it is clear these observations also have the
lowest S/N.   

\begin{figure*}
\centering
\includegraphics[angle=270, width=8cm]{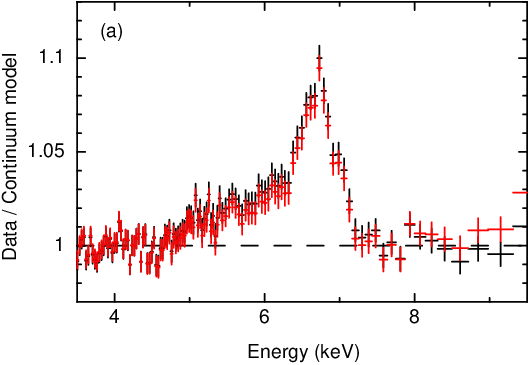}
\includegraphics[angle=270, width=8cm]{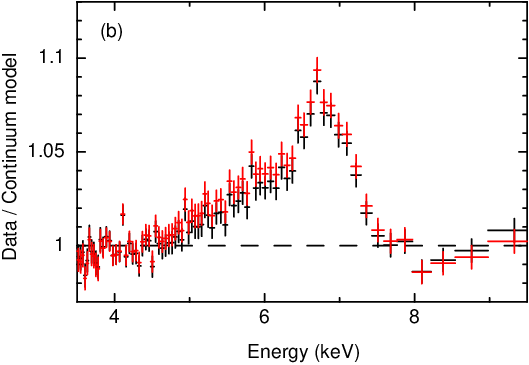}
\includegraphics[angle=270, width=8cm]{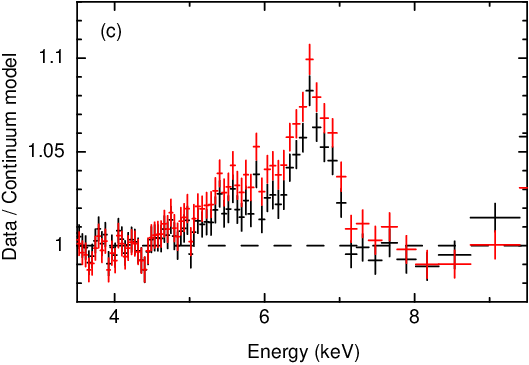}
\includegraphics[angle=270, width=8cm]{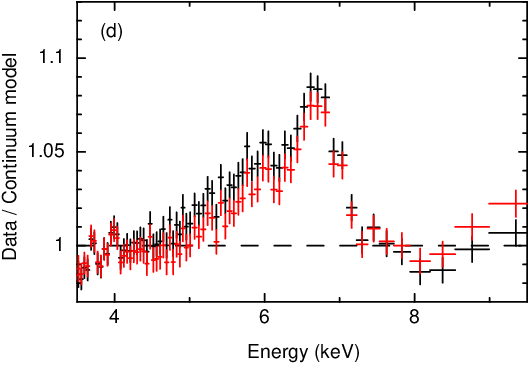}
\caption{A comparison of iron emission line profiles when fitting different
continuum models to the {\it Suzaku} data. Panel (a)
shows Serpens X-1, (b) GX~349+2 (observation 1), (c) GX~17+2 (observation 2),
and (d) 4U~1705$-$44 (observation 2). In all
panels, the spectra are from the combined front-illuminated XIS detectors. 
Black is when fitting with a diskbb component, while red is when the comptt
component is used instead.  Clearly, there is not a strong dependence on the
choice of continuum model tested here.}
\label{fig:comparison_diffcont}
\end{figure*}

It is interesting to note that the \suz\ and \xte\ observations almost all give consistent results with differing models.  While \xte\ has the
lowest spectral resolution, it has the highest effective area, whereas \suz\
has by the far the most superior spectral resolution.  We can try and
understand why there is a difference in the Gaussian line width with
different models for \exo\ and \sax\ by considering the line profiles seen by
\suz.  Firstly, in Figure~\ref{fig:comparison_diffcont} we show the \suz\ line
profiles (when fit by a diskline model) for both continuum models.  Clearly, the
line profiles vary very little with the different continuum models,
demonstrating that a robust line profile can be obtained with high spectral
resolution and also good S/N.  Note that the only \suz\ observation where there
is a large difference in $\sigma$ is the first observation of 4U~1705$-$44,
which is the observation with the lowest S/N of all the \suz\ observations, and one where in \citet{cackett10} we had to fix line parameters because they were poorly constrained.

\begin{figure}
\centering
\includegraphics[angle=270, width=8cm]{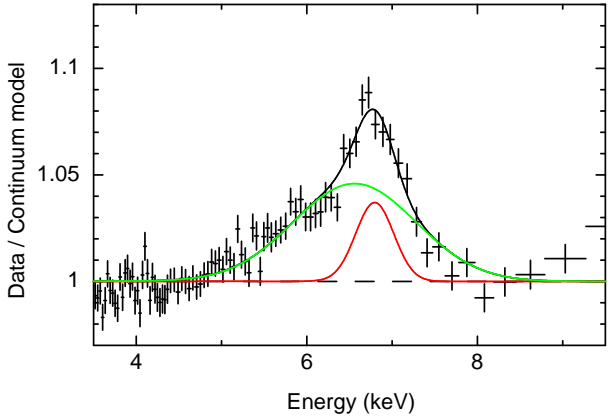}
\caption{A illustration that two Gaussians, one broad (green) and one narrower
(red), can fit the line profile well in the \suz{} data of GX 349+2, observation
1.  The combination of both Gaussians is shown as a black, solid line.}
\label{fig:twogauss}
\end{figure}

The \suz\  line profiles generally show a strong and quite narrow
line peak, with a broader, weaker red wing.  A single Gaussian does not fit the
line profile well and, in fact,  two Gaussians (one broad, one narrower)
generally provide a better fit \citep{cackett08}.  We also fit the
\suz\ data with two Gaussians, and give the results in
Table~\ref{tab:suz_2gauss} (for observations where the addition of a second
Gaussian improves the fit). Figure~\ref{fig:twogauss} illustrates how two
Gaussians can fit the \suz\ line profile of GX 349+2. 

\input{tab11.tex}

In the case of \xte-type spectral resolution, the details of the line shape are
blurred out by the line-spread function to a stage where the lines always
appear Gaussian.  The intermediate spectral resolution of \exo\ and \sax, can
show a hint of a red wing (see panel (c) in Fig.~\ref{fig:comparison}). However,
whether a broad or narrow Gaussian fits the line profile best can depend on the
underlying continuum model, though as we note above it is the observations with the largest uncertainty in line width that are the biggest outliers.

We therefore conclude that when different continuum models fit the data equally well, the line widths determined are generally consistent.

\section{Discussion}

We have studied spectra of four neutron star low-mass X-ray binaries
(Ser~X-1, GX~349+2, GX~17+2 and 4U~1705$-$44) from four different missions
(\exo, \sax, \xte, \suz) in order to compare the iron line profiles. Our main
result is that we find iron lines that show very similar line profiles between
the four different missions examined (Fig. \ref{fig:comparison}).  Lines in the
CCD spectra of \suz\ are not consistently broader than in the gas spectrometer
data, demonstrating that the broad profiles are not a consequence of pile-up or
other instrumental effects.  Moreover, we found three \sax\ observations (one of
GX 349+2, one of 4U~1705$-$44 and one of GX~17+2) that show evidence for
asymmetry, with a relativistic disk-line model providing a better fit than a
Gaussian.  However, for data of average quality, the spectral resolution of the
gas detectors ($\sim$0.5 keV for the best case looked at here, \sax/MECS) is not
good enough to clearly show asymmetric profiles.
 
We also found that generally the continuum model choice leads to consistent iron line profiles when the models fit equally well.  However, differences can arise and this is particularly a problem with the lower spectral
resolution of gas spectrometers.  However, with CCD spectral resolution and
high S/N, we demonstrated that the line profiles can be robustly determined,
regardless of the continuum model. This issue is more significant for neutron
star LMXBs than black hole X-ray binaries due to the typically higher
curvature and level of continuum degeneracy in the spectra of neutron star
LMXBs.

The extensive simulations of \citet{miller10}, the relativistic line
profile detected in pile-up-free observations in 4U 1728$-$34
\citep{egron11} in EPIC/pn ``timing'' mode, and the fact that the
\suz\ line profiles do not change regardless of the extraction region
used \citep{cackett10} already comprise a large body of evidence suggesting
that pile-up and instrumental effects are not the source of broad,
asymmetric lines seen in neutron star LMXBs. Furthermore, our analysis
of archival data presented here also shows that broad lines with
comparable widths as the \suz\ data are seen in gas spectrometer data,
where pile-up does not occur.  In addition, three \sax\ observations
also show evidence for asymmetric line profiles, with a relativistic
line model providing a better fit than a Gaussian, and with the diskline parameters comparable with those observed by \suz.

It is also important to note that inferred upper limits on stellar radii
are consistent with expectations from dense matter equations of state
\citep{cackett08}.  Furthermore, magnetic field estimates based on the inner
disk radius measured from the Fe lines in two pulsars, SAXJ1808.4$-$3658
\citep{cackett_j1808_09, papitto09} and IGR J17480-2446 in the globular cluster Terzan 5
\citep{miller11, papitto11}, are consistent with the magnetic field estimates determined
from timing methods.  Thus, the measured inner disk radii are consistent with
expectations.

The broadest iron lines in neutron star LMXBs, extend down to
approximately 4.5 keV, i.e. approximately a 2 keV redshift.  If due to Doppler
broadening alone, such a shift in emission would indicate high velocities of
the order $\sim$0.3c. This, of course, ignores any contribution from
gravitational redshifts and Comptonization which will also broaden the
line.  Certainly, in highly ionized disks, Comptonization will be an important
source of line broadening \citep[see e.g.][]{ross07}, but
self-consistent reflection modeling which include those effects also indicate
that the inner disk is close to the neutron star surface where relativistic
blurring is strong \citep{reis09_1705, cackett10,dai10}.

Reflection is further supported by the multiple observations of 4U~1705$-$44
considered by \citet{lin10}.  These authors show that the blackbody flux and
iron line flux are strongly correlated in the soft states.  This supports the
hypothesis that the blackbody flux (possibly originating from the boundary
layer) is the source of irradiating flux leading to reflection, at least in the
soft states.

Thus far, studies of iron line variability between the states has been
inconclusive.  In the majority of work that has looked at this issue, there is
no clear evidence of large changes between states
\citep{iaria09,cackett10,lin10,dai10}.  In the hard state
of 4U~1705$-$44 there has been tentative evidence that the line may be narrower
\citep{lin10,dai10}, though this may be an ionization effect
\citep{reis09_1705}. While a thorough study of line evolution with state is
beyond the scope of this paper, it is interesting to point out that we do seem
to see variability in the line profiles, at least in the case of GX~349+2.

\begin{figure}
\centering
\includegraphics[angle=270,width=8cm]{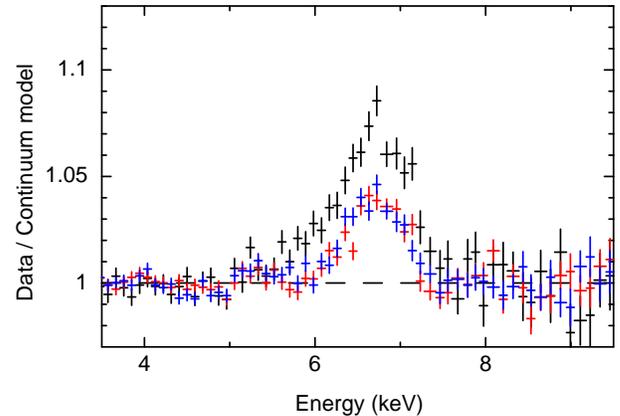}
\caption{A comparison of the line profiles from the three \sax\ observations of
GX 349+2.  Observation 1 (black) shows a stronger, broader line than
observations 2 (red) or 3 (blue).  See Table~\ref{tab:sax_bb_comptt_gauss} for
spectral parameters from these fits.}
\label{fig:gx349sax}
\end{figure}

The line profiles from the three \sax\ observations are shown in
Figure~\ref{fig:gx349sax}. Detailed analysis of these \sax\ observations have
previously been presented by \citet{disalvo01} and \citet{iaria04}, and both
those papers carefully looked at variability in the Fe line.  The variability we
see is consistent with their work.  \citet{disalvo01}
found a difference in the line EW when comparing spectra from flaring and
non-flaring periods of the first \sax\ observation -- the EW was larger in the
non-flaring spectrum when there is a hard power-law tail extending up to 100
keV.  \citet{iaria04} considered all the \sax\ observations, creating spectra
based on the location in the color-color diagram.  They found a very clear
trend showing that the line EW decreases with increasing source luminosity. 
They note that the decrease in EW is due to both an increase in the
continuum flux and a decrease in the line intensity.  Interestingly, the
spectra where the EW is strongest corresponds to the intervals where the hard
power-law tail is strongest.  Spectra from the flaring branch show the weakest
line, and that is also where the hard power-law tail is not detected.

\citet{cackett10} and \citet{dai10} both discuss reflection in these neutron
star LMXB systems, suggesting that, at least in the soft state, the irradiating
source of hard X-ray flux is the boundary layer.  However, in the first \sax\
observation of GX~349+2, there is also an additional hard power-law component at
higher energies \citep{disalvo01}.  If this originates from a different region,
for example a disk corona, then there could potentially be two sources of
irradiation with different geometries leading to reflection.

\section{Conclusions}
To summarize our main findings:
\begin{itemize}
 \item The iron line profiles seen in gas-based and CCD-based spectra are
consistent.  This demonstrates that the broad profiles are intrinsic to the
lines and are not due to instrumental effects (such as pile-up).
 \item Several \sax\ observations of GX~349+2, GX~17+2 and 4U~1705$-$44 all show
evidence for asymmetric line profiles.  A relativistic diskline model fits
better than a Gaussian line model as demonstrated by the posterior predictive
p-value and bootstrapping methods.
 \item We also found that generally the continuum model choice leads to consistent iron line profiles when the models fit equally well.  However, differences can arise and this is particularly a problem with the lower spectral resolution of gas spectrometers, but that line profiles determined
by \suz{} are generally robust to the continuum choice.
\end{itemize}

\acknowledgements
We thank Erik Kuulkers for interesting discussions and helpful suggestions.  ACF thanks the Royal Society for support.  RCR is supported by NASA through the Einstein Fellowship Program, grant No. 
PF1-120087 and is a member of the Michigan Society of Fellows. 

\bibliographystyle{apj}
\bibliography{apj-jour,felines}

\end{document}

%% file: tab1.tex
\tabletypesize{\scriptsize}
\begin{deluxetable*}{lccccccc}
\tablecolumns{8}
\tablewidth{0pc}
\tablecaption{Details of observations analyzed}
\tablehead{Source  & Mission & Obs. No. & Obs. ID & Start date & Exp. time & Source state & References \\
                            &   &    &  & (dd/mm/yy) & (ksec) & & }
\startdata
Serpens~X-1 &  \exo & 1 & 60208 & 08/09/85 & 50.5 & Banana & 1, 2, 3, 4 \\
Serpens~X-1 & \sax & 1 & 20835002 & 05/09/99 & 15.7/32.0/15.4  &  Banana & 5 \\
Serpens~X-1 & \xte & 1 & 20072-01-01-000 & 18/07/97 & 12.8 & Banana & 6,7 \\
Serpens~X-1 & \xte & 2 & 20072-01-04-000 & 31/07/97 & 14.5  & Banana & 6,7 \\
Serpens~X-1 & \xte & 3 & 40426-01-01-000 & 05/09/99 & 15.4 & Banana & 6,7 \\
Serpens~X-1 & \suz & 1 & 401048010 & 24/10/06 & 18/29 & Banana & 8,9 \\
\hline
GX~349+2 & \exo & 1 & 34067 & 10/09/84 & 18.8 & NB/FB & 2, 10 \\
GX~349+2 & \exo & 2 & 59605 & 31/08/85 & 65.0 & NB/FB & 2, 10, 11\\
GX~349+2 & \exo & 3 & 59663 & 01/09/85 & 94.3 & NB/FB & 2, 10, 11\\
GX~349+2 & \sax & 1 & 21009001  & 10/03/00 & 15.9/44.9/22.5 & FB with NB/FB vertex & 12 \\
GX~349+2 & \sax & 2 & 21240002 & 12/02/01 & 15.6/76.1/39.1 & FB with NB/FB vertex & 13 \\
GX~349+2 & \sax & 3 & 212400021 & 17/02/01 & 22.7/83.4/41.0  & FB with NB/FB vertex & 13 \\
GX~349+2 & \xte & 1 & 30042-02-01-000 & 09/01/98 & 17.0 &  FB & 14\\
GX~349+2 & \xte & 2 & 30042-02-01-04 & 13/01/98 & 13.6 & FB & 14\\
GX~349+2 & \xte & 3 & 30042-02-02-060 & 24/01/98 & 15.2 & NB/FB vertex & 14\\
GX~349+2 & \suz & 1 & 400003010 & 14/03/06 & 8/20 & NB & 8, 9 \\
GX~349+2 & \suz & 2 & 400003020 & 19/03/06 & 8/24 & NB & 8, 9 \\
\hline
GX~17+2 & \exo & 1 & 33715 & 05/09/84 & 25.1 & HB & 2, 15\\
GX~17+2 & \exo & 2 & 33781 & 06/09/84 & 26.1 & HB/NB vertex & 2, 15 \\
GX~17+2 & \exo & 3 & 58809 & 20/08/85 & 68.9 & NB & 2, 15 \\
GX~17+2 & \exo & 4 & 60698 & 15/09/85 & 42.6 & HB & 2, 15 \\
GX~17+2 & \sax & 1 & 21057001 & 04/10/99 & 11.6/41.1/19.6 & HB/NB & 16 \\
GX~17+2 & \sax & 2 & 210570011 & 05/10/99 & 22.0/75.2/36.9 & HB/NB & 16 \\
GX~17+2 & \sax & 3 & 210570012 & 07/10/99 & 13.2/48.3/23.4 & HB/NB & 16 \\
GX~17+2 & \sax & 4 & 20261011 & 03/04/97 & 2.3/10.6/4.5 & HB & 17 \\
GX~17+2 & \sax & 5 & 20261012 & 21/04/97 & 1.4/6.4/2.6 & HB & 17 \\
GX~17+2 & \xte & 1 & 20053-03-02-010 & 07/02/97 & 13.7 & NB & 18 \\
GX~17+2 & \xte & 2 & 30040-03-02-010 & 18/11/98 & 13.6 & FB & 18 \\
GX~17+2 & \xte & 3 & 30040-03-02-011 & 18/11/98 & 16.2 & FB & 18 \\
GX~17+2 & \suz & 1 & 402050010 & 19/09/07 & 5/15 & NB & 9 \\
GX~17+2 & \suz & 2 & 402050020 & 27/09/07 & 6/18 & NB & 9 \\
\hline
4U~1705$-$44 & \sax & 1 & 21292001 & 20/08/00 & 20.6/43.5/20.1 & Soft & 19, 20 \\
4U~1705$-$44 & \sax & 2 & 21292002 & 03/10/00 & 15.6/47.5/20.1 & Hard & 20, 21 \\
4U~1705$-$44 & \xte & 1 &  20073-04-01-00 & 01/04/97 & 12.8 & Hard & \\
4U~1705$-$44 & \xte & 2 & 70038-04-01-01 & 31/08/02 & 12.6  & Soft & \\
4U~1705$-$44 & \xte & 3 & 90170-01-01-000  & 26/10/05 & 16.1 & Soft & \\
4U~1705$-$44 & \suz & 1 & 401046010 & 29/08/06 & 14/14 & Hard &  9, 21, 22\\
4U~1705$-$44 & \suz & 2 & 401046020 & 18/09/06 & 17/15 & Soft &  9, 21, 22\\
4U~1705$-$44 & \suz & 3 & 401046030 & 06/10/06 & 18/17 & Soft &  9, 21, 22
\enddata
\tablecomments{Obs. No. is the observation number that we use in order to identify the observations.  The \sax\ exposure times are given for the LECS, MECS and PDS (in that order), while the \suz\ exposure times are the good time exposure for the individual XIS detectors and the HXD/PIN good time (in that order). Source state abbreviations: NB - normal branch, HB - horizontal branch, FB - flaring branch.}
\tablerefs{ 
(1) \citealt{white86},
(2) \citealt{gottwald95},
(3) \citealt{schulz89},
(4) \citealt{seon02},
(5) \citealt{oosterbroek01},
(6) \citealt{muno02},
(7) \citealt{gladstone07},
(8) \citealt{cackett08},
(9) \citealt{cackett10},
(10) \citealt{kuulkers95},
(11) \citealt{ponman88},
(12) \citealt{disalvo01},
(13) \citealt{iaria04},
(14) \citealt{zhang98b},
(15) \citealt{kuulkers97},
(16) \citealt{disalvo00},
(17) \citealt{farinelli05},
(18) \citealt{homan02},
(19) \citealt{piraino07},
(20) \citealt{fiocchi07},
(21) \citealt{lin10},
(22) \citealt{reis09_1705}
}
\label{tab:obs}
\end{deluxetable*}

%% file: tab2.tex
\tabletypesize{\tiny}
\begin{deluxetable*}{lccccccccccc}
\tablecolumns{12}
\tablewidth{0pc}
\tablecaption{Spectral fits from \exo\ observations with the an absorbed diskbb + bbody + Gaussian model.}
\tablehead{Source  & Obs. No. & $N_{\rm H}$ & \multicolumn{2}{c}{Disk blackbody}  & \multicolumn{2}{c}{Blackbody} &  \multicolumn{4}{c}{Gaussian} & $\chi^2$ (dof) \\
& & $10^{22}$ cm$^{-2}$ & kT (keV) & norm & kT (keV) & norm &  E (keV) & $\sigma$ (keV) & norm & EW (eV) & 
}
\startdata
Serpens~X$-$1 & 1 & $0.40\pm0.15$ & $1.33\pm0.07$ & $61\pm13$ & $2.10\pm0.03$ & $(3.0\pm0.1)\times10^{-2}$ &  $6.43^{+0.07}_{-0.03}$ & $0.72\pm0.09$ & $(6.9\pm1.2)\times10^{-3}$ & $169\pm36$ & 192.0 (183)
\\
GX~349+2  & 1 & $0.25\pm0.06$ & $2.09\pm0.10$ & $44\pm7$ & $2.65\pm0.16$ & $(4.1\pm0.1)\times10^{-2}$ &  $6.68\pm0.05$ & $0.47\pm0.08$ & $(1.4\pm0.2)\times10^{-2}$ & $102^{+16}_{-9}$ & 127.9 (127)
\\
GX~349+2 & 2 & $0.60\pm0.04$ & $2.35\pm0.05$ & $29\pm2$ & $3.46\pm0.27$ & $(2.5\pm0.5)\times10^{-2}$ &  $6.40^{+0.08}$ & $1.09\pm0.09$ & $(3.6\pm0.4)\times10^{-2}$ & $221^{+32}_{-23}$ & 130.4 (115)
\\
GX~349+2 & 3 & $0.75\pm0.05$ & $2.44\pm0.03$ & $29\pm1$ & $4.39\pm0.45$ & $(1.9\pm0.1)\times10^{-2}$ &  $6.63\pm0.10$ & $0.85\pm0.15$ & $(2.5\pm0.4)\times10^{-2}$ & $155\pm16$ & 120.1 (115)
\\
GX~17+2 & 1 & 2.38 (fixed) & $1.40\pm0.03$ & $192\pm16$ & $2.57\pm0.02$ & $0.117\pm0.002$ & $6.58\pm0.10$ & $0.80\pm0.14$ & $(1.9\pm0.5)\times10^{-2}$ & $131\pm23$ & 139.6 (185)
\\
GX~17+2 & 2 & 2.38 (fixed) & $1.55\pm0.04$ & $151\pm14$ & $2.46\pm0.03$ & $(9.2\pm0.4)\times10^{-2}$ & $6.68\pm0.14$ & $0.80\pm0.26$ & $(1.2\pm0.6)\times10^{-2}$ & $81\pm20$ & 157.8 (186)
\\
GX 17+2 & 3 & 2.38 (fixed) & $1.67\pm0.03$ & $103\pm6$ & $2.54\pm0.04$ & $(5.0\pm0.2)\times10^{-2}$ & $6.74\pm0.08$ & $0.54\pm0.11$ & $(1.2\pm0.2)\times10^{-2}$ & $99\pm15$ & 97.7 (112)
\\
GX 17+2 & 4 & 2.38 (fixed) & $1.46\pm0.03$ & $144\pm11$ & $2.80\pm0.02$ & $0.116\pm0.001$ & $6.4^{+0.16}$ & $1.87\pm0.14$ & $(5.9\pm0.9)\times10^{-2}$ & $424^{+329}_{-278}$ & 202.0 (124)
\enddata
\label{tab:exo_diskbb_bb_gauss}
\end{deluxetable*}

%% file: tab3.tex
\tabletypesize{\tiny}
\begin{deluxetable*}{lccccccccccccc}
\tablecolumns{14}
\tablewidth{0pc}
\tablecaption{Spectral fits from \exo\ observations with the an absorbed bbody + comptt + Gaussian model.}
\tablehead{Source  & Obs. No. & $N_{\rm H}$ & \multicolumn{2}{c}{Blackbody} &  \multicolumn{4}{c}{Comptonized component} & \multicolumn{4}{c}{Gaussian} & $\chi^2$ (dof) \\
& & $10^{22}$ cm$^{-2}$ & kT (keV) & norm & $T_0$ (keV) & kT (keV) & $\tau$ & norm & E (keV) & $\sigma$ (keV) & norm & EW (eV) & 
}
\startdata
Serpens~X$-$1 & 1 & 0.40 (fixed) & $0.87_{-0.05}^{+0.19}$ & $(1.1\pm0.4)\times10^{-2}$ & $0.35_{-0.35}^{+0.11}$ & $2.22\pm0.06$ & $8.6_{-1.1}^{+0.6}$ & $0.57_{-0.04}^{+0.14}$ & $6.42^{+0.09}$ & $0.73\pm0.10$ & $(7.0\pm1.7)\times10^{-3}$ & $173^{+77}_{-53}$ & 191.6 (182)
\\
GX~349+2  & 1  & $0.58\pm0.32$ & $1.33\pm0.08$ & $(5.1\pm0.4)\times10^{-2}$ & $0.46\pm0.05$ & $2.47\pm0.06$ & $7.9\pm0.5$ & $1.20_{-0.05}^{+0.17}$ & $6.71\pm0.05$ & $0.39\pm0.08$ & $(1.2\pm0.2)\times10^{-2}$ & $89^{+47}_{-8}$ & 124.6 (125)
\\
GX~349+2 & 2 & $0.5 (fixed)$ & $1.44\pm0.02$ & $(9.1\pm0.3)\times10^{-2}$ & $0.33_{-0.32}^{+0.05}$ & $2.91\pm0.06$ & $6.8\pm0.2$ & $1.2_{-0.1}^{+13.7}$ & $6.84\pm0.05$ & $0.31\pm0.07$ & $(9.7\pm1.4)\times10^{-3}$ & $67\pm9$ & 101.1 (113)
\\
GX~349+2 & 3 & $0.54^{+0.55}_{-0.17}$ & $1.52\pm0.03$ & $(9.9\pm0.3)\times10^{-2}$ & $0.47\pm0.10$ & $3.15\pm0.12$ & $6.0\pm0.5$ & $0.9_{-0.2}^{+0.3}$ & $6.82\pm0.06$ & $0.34\pm0.07$ & $(1.0\pm0.2)\times10^{-2}$ & $64^{+17}_{-7}$ & 109.9 (114)
\\
GX~17+2 & 1 & 2.38 (fixed) & $1.11\pm0.09$ & $(2.5\pm0.6)\times10^{-2}$ & $0.43\pm0.09$ & $3.09\pm0.08$ & $6.0\pm0.3$ & $1.6\pm0.2$ & $6.67\pm0.12$ & $0.48\pm0.16$ & $(8.8\pm3.1)\times10^{-3}$ & $61\pm16$ & 120.9 (183)
\\
GX~17+2 & 2 & 2.38 (fixed) & $1.32\pm0.08$ & $(5.0\pm0.6)\times10^{-2}$ & $0.51\pm0.05$ & $2.96\pm0.14$ & $5.6\pm0.4$ & $1.5\pm0.1$ & $6.75\pm0.12$ & $0.28\pm0.23$ & $4.6_{-0.1}^{+0.4} \times10^{-3}$ & $31\pm12$ & 142.6 (184)
\\
GX 17+2 & 3 & 2.38 (fixed) & $1.19\pm0.02$ & $(5.8\pm0.2)\times10^{-2}$ & $0.25\pm0.06$ & $2.68\pm0.04$ & $6.1\pm0.1$ & $2.3\pm0.5$ & $6.79\pm0.06$ & $0.44\pm0.09$ & $(9.4\pm1.7)\times10^{-3}$ & $79^{+175}_{-1}$ & 100.8 (117)
\\
GX 17+2 & 4 & 2.38 (fixed) & $1.46\pm0.09$ & $(1.7\pm0.2)\times10^{-2}$ & $0.24\pm0.05$ & $3.34\pm0.05$ & $5.8\pm0.1$ & $2.2\pm0.3$ & $6.79\pm0.09$ & $0.41\pm0.11$ & $(7.2\pm1.5)\times10^{-3}$ & $55^{+248}_{-1}$ & 150.4 (129)

\enddata
\label{tab:exo_bb_comptt_gauss}
\end{deluxetable*}

%% file: tab4.tex
\tabletypesize{\tiny}
\begin{deluxetable*}{lccccccccccccc}
\tablecolumns{14}
\tablewidth{0pc}
\tablecaption{Spectral fits from \xte\ observations with the an absorbed diskbb + bbody + power-law + Gaussian model.}
\tablehead{Source  & Obs. No. & $N_{\rm H}$ & \multicolumn{2}{c}{Disk blackbody}  & \multicolumn{2}{c}{Blackbody} &  \multicolumn{2}{c}{Power-law} & \multicolumn{4}{c}{Gaussian} & $\chi^2$ (dof) \\
& & $10^{22}$ cm$^{-2}$ & kT (keV) & norm & kT (keV) & norm &  $\Gamma$ & norm & E (keV) & $\sigma$ (keV) & norm & EW (eV) & 
}
\startdata
Serpens~X$-$1 & 1 & 0.4 (fixed) & $1.54\pm0.03$ & $49\pm3$ & $2.37\pm0.02$ & $(2.3\pm0.1)\times10^{-2}$ & -- & -- & $6.51\pm0.06$ & $0.48\pm0.12$ & $(4.8\pm0.7)\times10^{-3}$ & $104\pm21$ & 27.0 (38)
\\
Serpens~X$-$1 & 2 & 0.4 (fixed) & $1.66\pm0.04$ & $50\pm4$ & $2.39\pm0.03$ & $(3.3\pm0.2)\times10^{-2}$ & -- & -- & $6.51\pm0.06$ & $0.64\pm0.11$ & $(9.6\pm1.5)\times10^{-3}$ & $138\pm18$ & 25.1 (38)
\\
Serpens~X$-$1 & 3 & 0.4 (fixed) & $1.71\pm0.04$ & $41\pm3$ & $2.41\pm0.03$ & $(2.9\pm0.2)\times10^{-2}$ & -- & -- & $6.51\pm0.06$ & $0.52\pm0.12$ & $(7.0\pm1.1)\times10^{-3}$ & $110\pm20$ & 18.1 (32)
\\
GX~349+2 & 1 & 0.5 (fixed) & $1.98\pm0.03$ & $69\pm3$ & $2.64\pm0.03$ & $(6.5\pm0.3)\times10^{-2}$ & -- & -- & $6.40^{+0.01}$ & $1.05\pm0.05$ & $(5.9\pm0.3)\times10^{-2}$ & $302\pm23$ & 46.2 (51)
\\
GX~349+2 & 2 & 0.5 (fixed) & $2.28\pm0.04$ & $46\pm3$ & $2.87\pm0.06$ & $(5.1\pm0.6)\times10^{-2}$ & -- & -- & $6.40^{+0.02}$ & $1.26\pm0.06$ & $(6.8\pm0.4)\times10^{-2}$ & $306\pm25$ & 53.4 (52)
\\
GX~349+2 & 3 & 0.5 (fixed) & $1.77\pm0.02$ & $87\pm4$ & $2.56\pm0.02$ & $(6.4\pm0.2)\times10^{-2}$ & -- & -- & $6.40^{+0.07}$ & $0.82\pm0.09$ & $(2.9\pm0.3)\times10^{-2}$ & $186\pm18$ & 37.1 (51)
\\
GX~17+2 & 1 & 2.38 (fixed) & $1.70\pm0.01$ & $130\pm4$ & $2.68\pm0.01$ & $(7.5\pm0.2)\times10^{-2}$ & -- & -- & $6.51\pm0.10$ & $0.64\pm0.13$ & $(1.5\pm0.3)\times10^{-2}$ & $86\pm14$ & 45.2 (51)
\\
GX~17+2 & 2 & 2.38 (fixed) & $2.05\pm0.03$ & $81\pm3$ & $2.76\pm0.02$ & $0.125\pm0.004$ & -- & -- & $6.46_{-0.06}^{+0.09}$ & $0.85\pm0.12$ & $(3.9\pm0.6)\times10^{-2}$ & $135\pm16$ & 38.5 (52)
\\
GX~17+2 & 3 & 2.38 (fixed) & $1.88\pm0.02$ & $105\pm4$ & $2.64\pm0.02$ & $(9.7\pm0.3)\times10^{-2}$ & -- & -- & $6.40^{+0.05}$ & $0.92\pm0.07$ & $(5.4\pm0.4)\times10^{-2}$ & $218\pm19$ & 49.0 (52)
\\
4U~1705$-$44 & 1 & 0.8 (fixed) &  $2.28\pm0.05$ & $2.7\pm0.2$ & $4.68\pm0.01$ & $(1.2\pm0.2)\times10^{-2}$ & $0.6\pm0.8$ & $(2.1^{+9.0}_{-1.0})\times10^{-3}$ & $6.40^{+0.01}$ & $0.62\pm0.07$ & $(3.0\pm0.2)\times10^{-3}$ & $197\pm49$ &  65.2 (75)
\\
4U~1705$-$44 & 2 & 0.8 (fixed) & $1.99\pm0.04$ & $18\pm1$ & $2.62\pm0.05$ & $(1.62\pm0.01)\times10^{-2}$ & -- & -- & $6.40^{+0.02}$ & $0.57\pm0.09$ & $(6.7\pm0.7)\times10^{-3}$ & $130\pm16$ & 30.6 (42)
\\
4U~1705$-$44 & 3 & 0.8 (fixed) & $1.52\pm0.04$ & $19\pm2$ & $2.27\pm0.02$ & $(1.18\pm0.05)\times10^{-2}$ & -- & -- & $6.52\pm0.07$ & $0.32\pm0.15$ & $(1.6\pm0.3)\times10^{-3}$ & $82\pm26$ & 33.8 (42)
\enddata
\label{tab:rxte_diskbb_bb_gauss}
\end{deluxetable*}

%% file: tab5.tex
\tabletypesize{\tiny}
\begin{deluxetable*}{lccccccccccccc}
\tablecolumns{14}
\tablewidth{0pc}
\tablecaption{Spectral fits from \xte\ observations with the an absorbed bbody + comptt + Gaussian model.}
\tablehead{Source  & Obs. ID & $N_{\rm H}$ & \multicolumn{2}{c}{Blackbody} &  \multicolumn{4}{c}{Comptonized component} & \multicolumn{4}{c}{Gaussian} & $\chi^2$ (dof) \\
& & $10^{22}$ cm$^{-2}$ & kT (keV) & norm & $T_0$ (keV) & kT (keV) & $\tau$ & norm & E (keV) & $\sigma$ (keV) & norm & EW (eV) & 
}
\startdata
Serpens~X$-$1 & 1 & 0.4 (fixed) & $1.14\pm0.07$ & $(1.3\pm0.3)\times10^{-2}$ & $0.41_{-0.40}^{+0.14}$ & $2.58\pm0.04$ & $6.0\pm0.3$ & $0.72_{-0.22}^{+28.08}$ & $6.45_{-0.05}^{+0.08}$ & $0.53\pm0.15$ & $(5.2\pm1.4)\times10^{-2}$ & $110^{+139}_{-21}$ & 22.0 (36)
\\
Serpens~X$-$1 & 2 & 0.4 (fixed) & $1.21\pm0.08$ & $(2.1\pm0.4)\times10^{-2}$ & $0.44_{-0.43}^{+0.15}$ & $2.51\pm0.04$ & $6.8\pm0.3$  & $0.83_{-0.09}^{+20.36}$ & $6.45^{+0.09}_{-0.05}$ & $0.66\pm0.14$ & $(9.9\pm2.7)\times10^{-3}$ & $138^{+94}_{-24}$ & 21.6 (36)
\\
Serpens~X$-$1 & 3 & 0.4 (fixed) & $1.23\pm0.06$ & $(2.0\pm0.2)\times10^{-2}$ & $0.39_{-0.38}^{+0.25}$ & $2.50_{-0.03}^{+0.11}$ & $6.8\pm0.3$ & $0.82_{-0.17}^{13.80}$ & $6.47\pm0.07$ & $0.53\pm0.17$ & $(7.0_{-1.0}^{+3.7})\times10^{-3}$ & $106^{+6}_{-56}$ & 15.7 (30)
\\
GX~349+2 & 1 & 0.5 (fixed) & $1.36_{-0.06}^{+0.10}$ & $(7.5\pm2.3)\times10^{-2}$ & $0.59_{-0.58}^{+0.10}$ & $2.64\pm0.05$ & $6.8\pm0.8$ & $1.4_{-0.1}^{+13.8}$ & $6.40^{+0.20}$ & $0.99_{-0.15}^{+0.09}$ & $(4.9_{-2.0}^{+01.0})\times10^{-2}$ & $246_{-65}^{+94}$ & 36.8 (49)
\\
GX~349+2 & 2 & 0.5 (fixed) & $1.42\pm0.03$ & $0.11\pm0.01$ & $0.40\pm0.37$ & $2.67\pm0.03$ & $7.6\pm0.4$ & $1.6_{-0.2}^{20.4}$ & $6.58_{-0.18}^{+0.13}$ & $0.90\pm0.28$ & $(3.0\pm1.3)\times10^{-2}$ & $139^{+281}_{-31}$ & 32.4 (43)
\\
GX~349+2 & 3 & 0.5 (fixed) & $1.25_{-0.04}^{+0.20}$ & $(5.6\pm2.1)\times10^{-2}$ & $0.56_{-0.55}^{+0.14}$ & $2.65\pm0.06$ & $6.6\pm0.7$ & $1.31_{-0.06}^{+17.84}$ & $6.50_{-0.03}^{+0.12}$ & $0.73\pm0.20$ & $(2.5\pm1.0)\times10^{-2}$ & $162\pm45$ & 34.3 (49)
\\
GX~17+2 & 1 & 2.38 (fixed) & $1.83\pm0.25$ & $(1.9\pm0.8)\times10^{-2}$ & $0.73\pm0.03$ & $3.1\pm0.2$ & $4.5\pm0.5$ & $1.6\pm0.1$ & $6.40^{+0.04}$ & $1.03\pm0.09$ & $(4.1\pm0.7)\times10^{-2}$ & $228^{+74}_{-47}$ & 28.9 (49)
\\
GX~17+2 & 2 &  2.38 (fixed) & $1.91\pm0.18$ & $(6.1\pm1.3)\times10^{-2}$ & $0.72\pm0.04$ & $3.0\pm0.1$ & $5.5\pm0.6$ & $2.0\pm0.2$ & $6.40^{+0.07}$ & $1.10\pm0.10$ & $(7.1\pm1.3)\times10^{-2}$ & $248^{+73}_{-52}$ & 34.3 (50)
\\
GX~17+2 & 3 & 2.38 (fixed) & $1.79\pm0.24$ & $(4.3\pm0.9)\times10^{-2}$ & $0.75\pm0.04$ & $2.9\pm0.1$ & $5.2\pm0.6$ & $2.0\pm0.1$ & $6.40^{+0.05}$ & $1.06\pm0.08$ & $(7.7\pm1.0)\times10^{-2}$ & $318^{+72}_{-52}$ & 42.4 (50)
\\
4U~1705$-$44 & 1 & 0.8 (fixed) & $1.72\pm0.10$ & $(1.7\pm0.6)\times10^{-3}$ & $0.80\pm0.03$ & $10.7\pm0.9$ & $2.9\pm0.2$ & $(2.8\pm0.2)\times10^{-2}$ & $6.4^{+0.02}$ & $0.69\pm0.08$ & $(3.4\pm0.4)\times10^{-3}$ & $229\pm38$ & 66.1 (75)
\\
4U~1705$-$44 & 2 & 0.8 (fixed) & $2.87_{-0.07}^{+0.78}$ & $(9.6_{-6.7}^{+0.9})\times10^{-3}$ & $0.73\pm0.02$ & $2.0^{+0.3}$ & $7.6\pm0.3$ & $0.62^{+0.01}_{-0.06}$ & $6.40^{+0.06}$ & $0.67\pm0.10$ & $(8.0\pm1.0)\times10^{-3}$ & $158\pm30$ & 27.4 (40)
\\
4U~1705$-$44 & 3 & 0.8 (fixed) & $1.17\pm0.08$ & $(3.5\pm0.7)\times10^{-3}$ & $0.32_{-0.31}^{+0.14}$ & $2.4\pm0.1$ & $6.8\pm0.3$ & $0.35_{-0.09}^{+5.03}$ & $6.45_{-0.05}^{+0.07}$ & $0.34\pm0.15$ & $(1.6\pm0.4)\times10^{-3}$ & $82_{-6}^{+5559}$ & 23.0 (40) 

\enddata 
\label{tab:rxte_bb_comptt_gauss}
\end{deluxetable*}

%% file: tab6.tex
\tabletypesize{\tiny}
\begin{deluxetable*}{lccccccccccccc}
\tablecolumns{14}
\tablewidth{0pc}
\tablecaption{Spectral fits from \sax\ observations with the an absorbed diskbb + bbody + power-law + Gaussian model.}
\tablehead{Source  & Obs. & $N_{\rm H}$ & \multicolumn{2}{c}{Disk blackbody}  & \multicolumn{2}{c}{Blackbody} &  \multicolumn{2}{c}{Power-law} & \multicolumn{4}{c}{Gaussian} & $\chi^2$ (dof) \\
& & $10^{22}$ cm$^{-2}$ & kT (keV) & norm & kT (keV) & norm &  $\Gamma$ & norm & E (keV) & $\sigma$ (keV) & norm & EW (eV) & 
}
\startdata
Serpens~X$-$1 & 1  & $0.62\pm0.01$ & $1.46\pm0.02$ & $51\pm3$ & $2.21\pm0.04$ & $(3.0\pm0.1)\times10^{-2}$ & $2.98\pm0.06$ & $0.90\pm0.05$ & $6.40^{+0.01}$ & $0.96\pm0.08$ & $(1.2\pm0.2)\times10^{-2}$ & $234\pm31$ & 222.2 (133)
\\
GX~349+2 & 1  & \multicolumn{3}{l}{No fit with reduced $\chi^2 < 2$}
\\
GX~349+2 & 2  & \multicolumn{3}{l}{No fit with reduced $\chi^2 < 2$}
\\
GX~349+2 & 3  & \multicolumn{3}{l}{No fit with reduced $\chi^2 < 2$}
\\
GX~17+2 & 1  & $2.12\pm0.04$ & $1.81\pm0.03$ & $77\pm5$ & $2.79\pm0.02$ & $(7.0\pm0.3)\times10^{-2}$ & $2.4\pm0.2$ & $0.26_{-0.17}^{+0.32}$ & $6.60\pm0.06$ & $0.53\pm0.15$ & $(9.1\pm2.2)\times10^{-3}$ & $60\pm8$ & 203.1 (119)
\\
GX~17+2 & 2  & $2.12\pm0.05$ & $1.87\pm0.01$ & $70\pm2$ & $2.79\pm0.02$ & $(5.9\pm0.2)\times10^{-2}$ & $2.7\pm0.3$ & $0.27_{-0.19}^{+0.23}$ & $6.70\pm0.03$ & $0.33\pm0.05$ & $(5.8\pm0.6)\times10^{-3}$ & $40\pm4$ & 231.5 (119)
\\
GX~17+2 & 3  & \multicolumn{3}{l}{No fit with reduced $\chi^2 < 2$}
\\
GX~17+2 & 4  & $2.27\pm0.10$ & $1.70\pm0.06$ & $74\pm11$ & $3.01\pm0.02$ & $0.104\pm0.005$ & $2.6\pm0.1$ & $1.6\pm0.6$ & $6.83\pm0.12$ & $0.97\pm0.20$ & $(2.0\pm0.6)\times10^{-2}$ & $154\pm48$ & 175.7 (120)
\\
GX~17+2 & 5 & $2.08\pm0.10$ & $1.72\pm0.06$ & $86\pm12$ & $2.88\pm0.04$ & $(9.8\pm0.6)\times10^{-2}$ & $2.3\pm0.6$ & $0.3_{-0.3}^{+0.8}$ & $6.50^{+0.16}_{-0.10}$ & $0.72\pm0.26$ & $(1.3\pm0.6)\times10^{-2}$ & $84\pm29$ & 143.7 (120)
\\
4U 1705$-$44 & 1  & $1.72\pm0.07$ & $1.52\pm0.02$ & $42\pm3$ & $2.32\pm0.03$ & $(2.9\pm0.1)\times10^{-2}$ & $2.8\pm0.1$ & $0.6\pm0.2$ & $6.4^{+0.28}$ & $0.80\pm0.05$ & $(9.7\pm0.9)\times10^{-3}$ & $191\pm27$ & 162.6 (91)
\\
4U 1705$-$44 & 2  & $1.48\pm0.04$ & $2.03\pm0.07$ & $2.3\pm0.3$ & $4.4\pm0.2$ & $(5\pm1)\times10^{-3}$ & $0.6_{-0.5}^{+0.2}$* & $(1.1_{-0.8}^{+1.3})\times10^{-2}$ & $6.60\pm0.11$ & $0.76\pm0.24$ & $(1.4\pm0.5)\times10^{-3}$ & $145\pm88$ & 116.2 (106)

\enddata
\tablecomments{* The second observation of 4U 1705$-$44 has a strong hard tail extending to 100 keV.  A single power-law does not fit this tail well, thus we use a cut-off power-law instead.  We find the cut-off energy, $E_{\rm cut} = 28\pm5$ keV.}
\label{tab:sax_diskbb_bb_gauss}
\end{deluxetable*}

%% file: tab7.tex
\tabletypesize{\tiny}
\begin{deluxetable*}{lccccccccccccccc}
\tablecolumns{14}
\tablewidth{0pc}
\tablecaption{Spectral fits from \sax\ observations with the an absorbed bbody + comptt + power-law + Gaussian model.}
\tablehead{Source  & Obs & $N_{\rm H}$ & \multicolumn{2}{c}{Blackbody} &  \multicolumn{4}{c}{Comptonized component} & \multicolumn{2}{c}{Power-law} & \multicolumn{4}{c}{Gaussian} & $\chi^2$ (dof) \\
& & $10^{22}$ cm$^{-2}$ & kT (keV) & norm & $T_0$ (keV) & kT (keV) & $\tau$ & norm & $\Gamma$ & norm & E (keV) & $\sigma$ (keV) & norm & EW (eV) & 
}
\startdata
Serpens~X$-$1 & 1  & $0.56\pm0.05$ & $0.99\pm0.02$ & $(1.4\pm0.1)\times10^{-2}$ & $0.10\pm0.01$ & $2.58\pm0.04$ & $6.7\pm0.1$ & $2.0\pm0.2$ &  -- & -- & $6.40^{+0.26}$ & $1.01\pm0.07$ & $(1.5\pm0.2)\times10^{-2}$ & $288_{-26}^{+40}$ & 226.6 (132) 
\\
GX~349+2 & 1  & $0.52\pm0.03$ & $1.44\pm0.02$ & $(6.2\pm0.3)\times10^{-2}$ & $0.45\pm0.01$ & $2.63\pm0.04$ & $6.7\pm0.2$ & $1.18\pm0.03$ & $2.1\pm0.3$ & $(4.8_{-3.4}^{+14.2})\times10^{-2}$ & $6.68\pm0.03$ & $0.40\pm0.04$ & $(1.1\pm0.1)\times10^{-2}$ & $80\pm7$ & 160.7 (105)
\\
GX~349+2 & 2  & $0.52\pm0.01$ & $0.60\pm0.01$ & $(6.2\pm0.1)\times10^{-2}$ & $1.39\pm0.01$ & $2.94\pm0.02$ & $4.2\pm0.1$ & $1.20\pm0.01$ & -- & -- & $6.76\pm0.02$ & $0.20\pm0.03$ & $(5.9\pm0.4)\times10^{-3}$ & $31\pm3$ & 182.9 (100)
\\
GX~349+2 & 3  & $0.52\pm0.01$ & $0.62\pm0.01$ & $(6.0\pm0.1)\times10^{-2}$ & $1.43\pm0.01$ & $3.08\pm0.04$ & $3.6\pm0.1$ & $1.00\pm0.02$ & -- & -- & $6.71\pm0.02$ & $0.24\pm0.03$ & $(5.9\pm0.4)\times10^{-3}$ & $34\pm3$ & 189.2 (116)
\\
GX~17+2 & 1  & $2.27\pm0.09$ & $1.50\pm0.05$ & $(6.6\pm0.4)\times10^{-2}$ & $0.57\pm0.01$ & $3.10\pm0.07$ & $5.5\pm0.3$ & $0.99\pm0.03$ & $3.1\pm0.1$ & $2.6\pm0.7$ & $6.68\pm0.04$ & $0.21\pm0.05$ & $(4.0\pm0.6)\times10^{-3}$ & $27\pm6$ & 146.7 (117)
\\
GX~17+2 & 2  & $2.33\pm0.10$ & $1.44\pm0.03$ & $(6.8\pm0.2)\times10^{-2}$ & $0.54\pm0.01$ & $3.01\pm0.06$ & $5.6\pm0.2$ & $1.06\pm0.02$ & $3.5\pm0.2$ & $3.0\pm0.8$ & $6.73\pm0.03$ & $0.17\pm0.05$ & $(3.3\pm0.4)\times10^{-3}$ & $23\pm4$ & 162.6 (117)
\\
GX~17+2 & 3  & $2.18\pm0.10$ & $1.54\pm0.03$ & $(6.6\pm0.2)\times10^{-2}$ & $0.56\pm0.01$ & $3.18\pm0.05$ & $5.5\pm0.1$ & $1.02\pm0.02$ & $3.2\pm0.1$ & $2.0\pm0.6$ & $6.73\pm0.05$ & $0.25\pm0.08$ & $(3.1\pm0.6)\times10^{-3}$ & $21\pm6$ & 174.7 (118)
\\
GX~17+2 & 4 & $2.25\pm0.17$ & $1.59\pm0.08$ & $(3.2\pm0.2)\times10^{-2}$ & $0.56\pm0.03$ & $3.37\pm0.06$ & $6.0\pm0.2$ & $0.89\pm0.05$ & $2.8\pm0.1$ & $2.6\pm1.1$ & $6.77\pm0.06$ & $0.20\pm0.10$ & $(4.3\pm1.0)\times10^{-3}$ & $32\pm11$ & 150.1 (118)
\\
GX~17+2 & 5  & $2.10\pm0.29$ & $1.57\pm0.11$ & $(5.3\pm0.7)\times10^{-2}$ & $0.57\pm0.03$ & $3.25\pm0.14$ & $5.6\pm0.4$ & $0.99\pm0.06$ & $3.0\pm0.3$ & $1.8\pm1.5$ & $6.68\pm0.10$ & $0.17\pm0.17$ & $(3.3\pm1.5)\times10^{-3}$ & $22\pm15$ & 132.2 (118)
\\
4U 1705$-$44 & 1  & $1.57\pm0.15$ & $1.51\pm0.03$ & $(2.1\pm0.2)\times10^{-2}$ & $0.44\pm0.02$ & $2.69\pm0.06$ & $5.9\pm0.2$ & $0.49\pm0.03$ & $2.8\pm0.2$ & $0.49\pm0.33$ & $6.56\pm0.04$ & $0.27\pm0.06$ & $(2.9\pm0.4)\times10^{-3}$ & $58\pm14$ & 141.9 (89)
\\
4U 1705$-$44 & 2  & $0.91_{-0.02}^{+0.28}$ & -- & -- & $0.73\pm0.01$ & $6.65\pm0.52$ & $4.0\pm0.1$ & $(3.2\pm0.2)\times10^{-2}$ & $-0.74^{+1.8}_{-0.8}$ * & $(1.0^{+27.2}_{-0.9})\times10^{-3}$ & $6.4^{+0.4}$ & $1.12\pm0.09$ & $(3.5\pm0.5)\times10^{-3}$ & $375_{-18}^{+81}$ & 115.2 (106)
\enddata 
\tablecomments{* The strong hard tail extending to 100 keV requires a cut-off power-law.  We find the cut-off energy, $E_{\rm cut} = 20_{-3}^{+12}$ keV.}
\label{tab:sax_bb_comptt_gauss}
\end{deluxetable*}

%% file: tab8.tex
\tabletypesize{\tiny}
\begin{deluxetable*}{lccccccccccccc}
\tablecolumns{14}
\tablewidth{0pc}
\tablecaption{Spectral fits from \suz\ observations with the an absorbed diskbb + bbody + power-law + Gaussian model.}
\tablehead{Source  & Obs. & $N_{\rm H}$ & \multicolumn{2}{c}{Disk blackbody}  & \multicolumn{2}{c}{Blackbody} &  \multicolumn{2}{c}{Power-law} & \multicolumn{4}{c}{Gaussian} & $\chi^2$ (dof) \\
& & $10^{22}$ cm$^{-2}$ & kT (keV) & norm & kT (keV) & norm &  $\Gamma$ & norm & E (keV) & $\sigma$ (keV) & norm & EW (eV) & 
}
\startdata
Serpens~X$-$1 & 1  & $0.60\pm0.01$ & $1.24\pm0.01$ & $91\pm3$ & $2.17\pm0.02$ & $(5.1\pm0.1)\times10^{-2}$ & $2.7\pm0.1$ & $0.65\pm0.04$ & $6.65\pm0.01$ & $0.26\pm0.02$ & $(2.5\pm0.2)\times10^{-3}$ & $44\pm3$ & 1567.7 (1107)
\\
GX~349+2 & 1  & $0.79\pm0.02$ & $1.60\pm0.03$ & $95\pm6$ & $2.25\pm0.02$ & $0.117\pm0.003$ & $2.1\pm0.2$ & $0.17\pm0.11$ & $6.55\pm0.03$ & $0.58\pm0.04$ & $(1.4\pm0.1)\times10^{-2}$ & $87\pm6$ & 1341.9 (1108)
\\
GX~349+2 & 2  & $0.88\pm0.02$ & $1.50\pm0.02$ & $97\pm4$ & $2.30\pm0.02$ & $(9.5\pm0.1)\times10^{-2}$ & $2.4\pm0.1$ & $0.44\pm0.09$ & $6.71\pm0.02$ & $0.35\pm0.03$ & $(7.5\pm0.5)\times10^{-3}$ & $62\pm4$ & 1293.6 (1108)
\\
GX~17+2 & 1  & $2.24\pm0.11$ & $1.77\pm0.02$ & $83\pm4$ & $2.68\pm0.03$ & $(5.9\pm0.2)\times10^{-2}$ & $2.9\pm0.4$ & $0.84\pm0.68$ & $6.57\pm0.05$  & $0.28\pm0.05$ & $(4.3\pm0.7)\times10^{-3}$ & $30\pm6$ & 644.0 (608)
\\
GX~17+2 & 2  & $2.27\pm0.07$ & $1.88\pm0.02$ & $71\pm3$ & $2.64\pm0.02$ & $(7.6\pm0.3)\times10^{-2}$ & $3.0\pm0.2$ & $1.1\pm0.5$ & $6.60\pm0.02$ & $0.21\pm0.02$ & $(5.4\pm0.5)\times10^{-3}$ & $32\pm4$ & 665.9 (608)
\\
4U~1705$-$44 & 1  & 2.0 (fixed) & $(8.4\pm0.8)\times10^{-2}$ & $(7.7^{+17.2}_{-5.8})\times10^{7}$ & $1.18\pm0.06$ & $(6.6\pm0.5)\times10^{-4}$ & $1.67\pm0.01$ & $0.127\pm0.02$ & $6.55\pm0.03$ & $0.20\pm0.03$ & $(2.5\pm0.4)\times10^{-4}$ & $41\pm11$ & 961.8 (1123)
\\
4U~1705$-$44 & 2  & $2.01\pm0.02$ & $1.23\pm0.02$ & $44\pm3$ & $2.18\pm0.03$ & $(2.3\pm0.1)\times10^{-2}$ & $3.0\pm0.1$ & $0.79\pm0.05$ & $6.4^{+0.1}$ & $0.68\pm0.03$ & $(3.5\pm0.3)\times10^{-3}$ & $121\pm16$ & 1364.5 (1123)
\\
4U~1705$-$44 & 3  & $1.98\pm0.03$ & $0.74\pm0.01$ & $139\pm15$ & $1.84\pm0.01$ & $(9.8\pm0.1)\times10^{-3}$ & $2.9\pm0.1$ & $0.38\pm0.05$ & $6.4^{+0.2}$ & $0.40\pm0.05$ & $(4.5\pm0.8)\times10^{-4}$ & $40\pm14$ & 1275.8 (1122)
\label{tab:suz_diskbb_bb_gauss}
\enddata
\end{deluxetable*}

%% file: tab9.tex
\tabletypesize{\tiny}
\begin{deluxetable*}{lccccccccccccccc}
\tablecolumns{14}
\tablewidth{0pc}
\tablecaption{Spectral fits from \suz\ observations with the an absorbed bbody + comptt + power-law + Gaussian model.}
\tablehead{Source  & Obs & $N_{\rm H}$ & \multicolumn{2}{c}{Blackbody} &  \multicolumn{4}{c}{Comptonized component} & \multicolumn{2}{c}{Power-law} & \multicolumn{4}{c}{Gaussian} & $\chi^2$ (dof) \\
& & $10^{22}$ cm$^{-2}$ & kT (keV) & norm & $T_0$ (keV) & kT (keV) & $\tau$ & norm & $\Gamma$ & norm & E (keV) & $\sigma$ (keV) & norm & EW (eV) & 
}
\startdata
Serpens~X$-$1 & 1  & $0.57\pm0.01$ & $0.75\pm0.01$ & $(8.6\pm0.4)\times10^{-3}$ & $0.15\pm0.01$ & $2.44\pm0.01$ & $7.8\pm0.1$ & $1.54\pm0.09$ & -- & -- & $6.65\pm0.01$ & $0.26\pm0.02$ & $(2.5\pm0.1)\times10^{-3}$ & $44\pm3$ & 1605.7 (1109)
\\
GX~349+2 & 1  & $0.55\pm0.01$ & $1.37\pm0.07$ & $(4.8\pm0.4)\times10^{-2}$ & $0.45\pm0.01$ & $2.52\pm0.04$ & $7.9\pm0.3$ & $1.48\pm0.04$ & -- & -- & $6.67\pm0.03$ & $0.38\pm0.05$ & $(8.6\pm1.1)\times10^{-3}$ & $54\pm5$ & 1476.2 (1108)
\\
GX~349+2 & 2  & $0.93\pm0.01$ & $0.94\pm0.01$ & $(3.00\pm0.08)\times10^{-2}$ & $0.13\pm0.01$ & $2.51\pm0.01$ & $8.31\pm0.06$ & $2.6\pm0.1$ & -- & -- & $6.65\pm0.03$ & $0.49\pm0.05$ & $(1.1\pm0.1)\times10^{-2}$ & $59\pm11$ & 1367.3 (1108)
\\
GX~17+2 & 1  & $2.55\pm0.06$ & $1.30\pm0.04$ & $(4.8\pm0.4)\times10^{-2}$ & $0.60\pm0.02$ & $2.83\pm0.05$ & $6.1\pm0.2$ & $1.10\pm0.02$ & $3.5\pm0.1$ & $5.1\pm0.7$ & $6.54\pm0.06$ & $0.31\pm0.07$ & $(4.6\pm1.0)\times10^{-3}$ & $32\pm7$ & 618.7 (607)
\\
GX~17+2 & 2  & $2.50\pm0.06$ & $1.36\pm0.05$ & $(5.3\pm0.3)\times10^{-2}$ & $0.56\pm0.02$ & $2.73\pm0.05$ & $6.8\pm0.3$ & $1.27\pm0.02$ & $3.4\pm0.1$ & $4.6\pm0.6$ & $6.60\pm0.02$ & $0.20\pm0.03$ & $(5.0\pm0.6)\times10^{-3}$ & $30\pm4$ & 637.6 (606)
\\
4U~1705$-$44 & 1  & $1.56\pm0.03$ & $0.96\pm0.01$ & $(2.23\pm0.02)\times10^{-3}$ & $0.12_{-0.01}^{+0.12}$ & $6.6\pm0.1$ & $5.94\pm0.03$ & $(3.5\pm0.1)\times10^{-2}$ & -- & -- & $6.4^{+0.01}$ & $0.87\pm0.05$ & $(1.4\pm0.1)\times10^{-2}$ & $239_{-22}^{+485}$ & 1009.2 (1123)
\\
4U~1705$-$44 & 2  & $1.90\pm0.01$ & $0.61\pm0.02$ & $(1.9\pm0.3)\times10^{-3}$ & $0.14\pm0.02$ & $2.54\pm0.01$ & $6.85\pm0.07$ & $1.0\pm0.1$ & -- & -- & $6.4^{+0.2}$ & $0.60\pm0.03$ & $(2.7\pm0.2)\times10^{-3}$ & $92\pm15$ & 1368.2 (1125)
\\
4U~1705$-$44 & 3  & $1.59\pm0.02$ & $1.98\pm0.03$ & $(7.1\pm0.4)\times10^{-3}$ & $0.36\pm0.01$ & $3.9\pm0.5$ & $3.1\pm0.4$ & $0.13\pm0.02$ & -- & -- & $6.4^{+0.1}$ & $0.46\pm0.05$ & $(6.8\pm1.0)\times10^{-4}$ & $62\pm19$ & 1261.6 (1123)
\enddata 
\label{tab:suz_bb_comptt_gauss}
\end{deluxetable*}

%% file: tab10.tex
\tabletypesize{\tiny}
\begin{deluxetable*}{lccccccccccccccccc}
\tablecolumns{16}
\tablewidth{0pc}
\tablecaption{Spectral fits to \sax\ observations using a relativistic emission line model.}
\tablehead{Source  & Obs & $N_{\rm H}$ & \multicolumn{2}{c}{Blackbody} &  \multicolumn{4}{c}{Comptonized component} & \multicolumn{2}{c}{Power-law} & \multicolumn{6}{c}{Diskline} & $\chi^2$ (dof) \\
& & $10^{22}$ cm$^{-2}$ & kT (keV) & norm & $T_0$ (keV) & kT (keV) & $\tau$ & norm & $\Gamma$ & norm & E (keV) & $\beta$ & $R_{\rm in}$ & $i (^\circ)$ & norm & EW (eV) & 
}
\startdata
GX~349+2 & 1  & $0.51\pm0.02$ & $1.45\pm0.02$ & $(6.0\pm0.2)\times10^{-2}$ & $0.47\pm0.01$ & $2.63\pm0.04$ & $6.7\pm0.2$ & $1.18\pm0.02$ & $2.1\pm0.2$ & $(5.4_{-3.2}^{+5.8})\times10^{-2}$ & $6.97_{-0.35}$ & $-3.0\pm0.4$ & $8.3_{-1.9}^{+1.0}$ & $22_{-2}^{+8}$ & $(1.2_{-0.1}^{+0.3})\times10^{-2}$ & $111\pm13$ & 138.2 (103)
\\
4U 1705$-$44 & 1  & $1.66\pm0.17$ & $1.58\pm0.07$ & $(1.8\pm0.3)\times10^{-2}$ & $0.48\pm0.03$ & $2.64\pm0.12$ & $6.0\pm0.5$ & $0.47\pm0.06$ & $2.9\pm0.1$ & $0.8\pm0.5$ & $6.65_{-0.25}^{+0.32}$ & $-5.3^{+1.6}_{-4.7}$ & $11_{-3}^{+8}$ & $29\pm6$ & $(5.6\pm0.6)\times10^{-3}$ & $120\pm20$ & 113.7 (87) 
\\
GX~17+2 & 1  & $2.33\pm0.09$ & $1.55\pm0.06$ & $(5.7\pm0.4)\times10^{-2}$ & $0.61\pm0.01$ & $3.11\pm0.08$ & $5.4\pm0.3$ & $1.00\pm0.04$ & $3.18\pm0.08$ & $3.2\pm0.7$ & $6.47^{+0.23}$ & $<-4.8$ & $18^{+5}_{-9}$ & $37\pm8$ & $(9.0\pm0.9)\times10^{-3}$ & $67^{+23}_{-32}$ & 132.0 (115) 
\enddata 
\tablecomments{$R_{\rm in}$ is given in GM/c$^2$. In the diskline model, the outer disk radius is fixed at $R_{\rm out} = 1000$ GM/c$^2$ in all cases.}
\label{tab:diskline}
\end{deluxetable*}

%% file: tab11.tex
\tabletypesize{\tiny}
\begin{deluxetable*}{lcccccccccc}
\tablecolumns{11}
\tablewidth{0pc}
\tablecaption{Gaussian line parameters when fitting two Gaussians to \suz\ observations with an absorbed diskbb + bbody + power-law continuum model.}
\tablehead{Source  & Obs. & \multicolumn{4}{c}{Gaussian} & \multicolumn{4}{c}{Gaussian} &  $\chi^2$ (dof) \\
& &  E (keV) & $\sigma$ (keV) & norm & EW (eV) &  E (keV) & $\sigma$ (keV) & norm & EW (eV) 
}
\startdata
Serpens~X$-$1 & 1 & $6.40^{+0.16}$ & $0.64\pm0.04$ & $(3.3\pm0.4)\times10^{-3}$ & $53\pm10$ & $6.70\pm0.01$ & $0.14\pm0.02$ & $(1.1\pm0.1)\times10^{-3}$ & $19\pm5$ & 1464.6 (1104)
\\
GX~349+2 & 1  & $6.40^{+0.22}$ & $0.72\pm0.05$ & $(1.4\pm0.1)\times10^{-2}$ & $80\pm10$ & $6.79\pm0.03$ & $0.21\pm0.03$ & $(3.1\pm0.5)\times10^{-3}$ & $20\pm8$ & 1279.0 (1105)
\\
GX~349+2 & 2  & $6.40^{+0.34}$ & $0.73\pm0.07$ & $(6.5\pm1.1)\times10^{-2}$ & $48\pm13$ & $6.77\pm0.02$ & $0.24\pm0.02$ & $(4.0\pm0.5)\times10^{-3}$ & $34\pm10$ & 1253.5 (1105)
\\
GX~17+2 & 2  & $6.40^{+0.29}$ & $0.50\pm0.08$ & $(4.9\pm1.2)\times10^{-2}$ & $27\pm11$ & $6.64\pm0.02$ & $0.14\pm0.03$ & $(2.9\pm0.6)\times10^{-3}$ &  $17\pm8$ & 650.9 (605)
\\
4U~1705$-$44 & 2  & $6.40^{+0.12}$ & $0.72\pm0.04$ & $(3.3\pm0.3)\times10^{-3}$ & $111\pm19$ & $6.70\pm0.03$ & $0.12\pm0.03$ & $(2.5\pm0.6)\times10^{-4}$ & $9\pm6$ & 1338.2 (1120)
\enddata
\label{tab:suz_2gauss}
\end{deluxetable*}